\documentclass[12pt, hidelinks]{article}
\usepackage{graphicx}
\usepackage[sorting=none, style=ieee]{biblatex}
\usepackage{setspace}
\usepackage[breaklinks=true]{hyperref}
\usepackage{microtype}
\usepackage[center]{caption}
\usepackage[margin=1.5in, tmargin=2.0in]{geometry}
\usepackage{epigraph}
\usepackage{titletoc}
\usepackage{tocloft}
\usepackage[usenames,dvipsnames]{color}
\usepackage[T1]{fontenc}
\usepackage{tgpagella}
\graphicspath{ {./figures/} }
\title{A Transdisciplinary Approach to Cybersecurity: A Framework for Encouraging Transdisciplinary Thinking}
\author{Emily Kesler}
\date{2024}

\usepackage{titlesec}
\titleformat*{\section}{\large\bfseries}
\titleformat*{\subsection}{\normalsize\bfseries}

\begin{document}

\begin{titlepage}
    \begin{center}
        \vspace*{1cm}

        \large
        \textbf{A TRANSDISCIPLINARY APPROACH TO CYBERSECURITY:}
        
        \textbf{A FRAMEWORK FOR ENCOURAGING \\ TRANSDISCIPLINARY THINKING}
            
        \normalsize
        \vspace{0.5cm}
        by
        
        \vspace{0.3cm}
        Emily Kesler
        \vfill
            
        Submitted in Partial Fulfillment \\
        of the Requirements for the Degree of \\
        Master of Science in Transdisciplinary Cybersecurity \\
        with Independent Study in Transdisciplinary Cybersecurity
            
        \vspace{0.3cm}

        \includegraphics[scale=0.6]{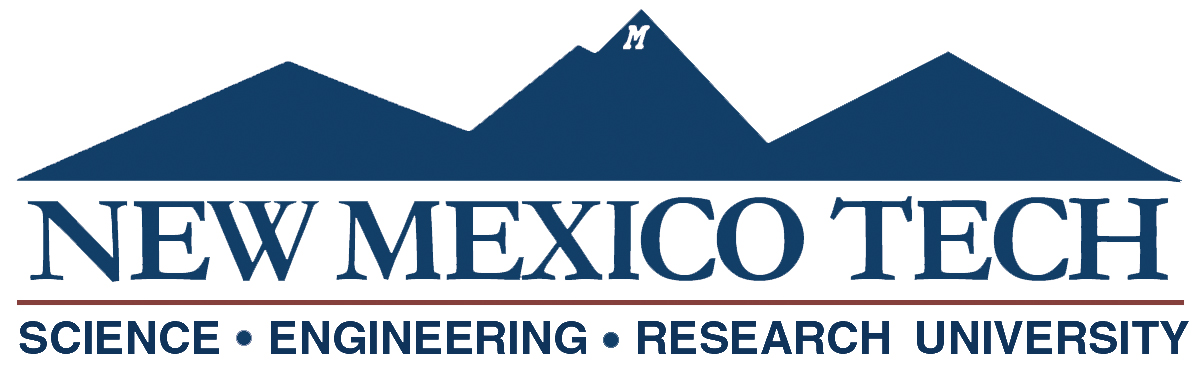}

        \vspace{0.3cm}
            
        New Mexico Institute of Mining and Technology \\
        Socorro, New Mexico \\
        May 2024
        
    \end{center}
\end{titlepage}

\newpage
\thispagestyle{empty}
\begin{center}
    \setlength\epigraphwidth{.8\textwidth}
    \setlength\epigraphrule{0pt}
    \renewcommand{\epigraphsize}{\normalsize}
    \renewcommand {\sourceflush} {center}
    \vspace*{\fill} 
    \begin{quote} 
    \centering 
    \epigraph{``See a need, fill a need.”}{---Bigweld, \textit{Robots} (2005)}
    \end{quote}
    \vspace*{\fill}

\end{center}

\newpage

\clearpage
\pagenumbering{roman} 
\thispagestyle{empty}
\begin{center}
    \large
    \textbf{ABSTRACT}
\end{center}

\indent 

Classical cybersecurity is often perceived as a rigid science discipline filled with computer scientists and mathematicians. However, due to the rapid pace of technology development and integration, new criminal enterprises, new defense tactics, and the understanding of the human element, cybersecurity is quickly beginning to encompass more than just computers. Cybersecurity experts must broaden their perspectives beyond traditional disciplinary boundaries to provide the best protection possible. They must start to practice transdisciplinary cybersecurity. Taking influence from the Stakeholder Theory in business ethics, this paper presents a framework to encourage transdisciplinary thinking and assist experts in tackling the new challenges of the modern day. The framework uses the simple Think, Plan, Do approach to enable experts to develop their transdisciplinary thinking. The framework is intended to be used \textit{as an evaluation tool for existing cybersecurity practices or postures}, \textit{as a development tool to engage with other disciplines to foster learning and create new methods}, and \textit{as a guidance tool to encourage new ways of thinking about, perceiving, and executing cybersecurity practices}. For each of those intended uses, a use case is presented as an example to showcase how the framework might be used. The ultimate goal of this paper is not the framework but transdisciplinary thinking. By using the tool presented here and developing their own transdisciplinary thinking, cybersecurity experts can be better prepared to face cybersecurity’s unique and complex challenges.

\vspace{0.5cm}

\noindent \textbf{Keywords: }cybersecurity; transdisciplinary; transdisciplinary cybersecurity; transdisciplinary thinking; framework; transdisciplinary framework

\newpage

\section*{\center ACKNOWLEDGMENTS}
\vspace{1pc}
\indent

It is with the utmost gratitude that I thank Dr. Lorie Liebrock, my academic advisor and research advisor, for her phenomenal patience, encouragement, feedback, guidance, dedication, and support throughout this project. I could not have completed my project or degree without her substantial contributions.

I want to express my deepest appreciation to Drs. Dongwan Shin and Stewart Thompson, my committee members, who generously offered their time, expertise, and support to help me shape, refine, and complete this project.

I am also grateful to Dr. Christopher ChoGlueck, whose Ethics of Cybersecurity course introduced me to Stakeholder Theory. His expert opinion on ethical challenges and his encouraging nature gave me a solid foundation on which to build this project.

I would like to extend my sincere thanks to Professor Christine Horwege for her enthusiasm, kindness, feedback, and willingness to help me throughout this project.

In addition, I’d like to offer my sincerest thanks to Susan Howard, who offered me her knowledge, feedback, and support throughout this project. Her expertise in Operational Technology and electric vehicles and her willingness to listen to and brainstorm with me have significantly improved my project and the resulting paper.

I want to give special thanks to my friend Emily Gray, who was my sounding board during the development phase of this project. Her social science background and expertise offered profound insight into my preliminary topic choices for this project. It was only after an incredibly engaging discussion with her that I decided to write about developing a transdisciplinary approach to cybersecurity.

Finally, I could not have undertaken this journey without my loving friends and family. Their support and assistance have been instrumental in sustaining me as I worked on this project and my degree.

\newpage

\thispagestyle{empty}
\renewcommand{\contentsname}{\hfill\bfseries\large TABLE OF CONTENTS \hfill \vspace{1pc}}
\renewcommand{\cftaftertoctitle}{\hfill}

\tableofcontents

\newpage

\thispagestyle{empty}
\renewcommand{\listfigurename}{\large \hfill LIST OF FIGURES \hfill \vspace{1pc}}
\renewcommand{\cftfigpresnum}{\bfseries Figure }
\settowidth{\cftfignumwidth}{Figure 00.0000}
\listoffigures

\newpage
\thispagestyle{empty}
\begin{center}
    \includegraphics[scale=0.6]{figures/NMTLogo.jpg}

    \normalsize
    
    \flushleft This Independent Study is accepted on behalf of the faculty of the Institute by the following committee:
    \vspace{0.5cm}
    
    \begin{tabular}[b]{@{} p{5.5in} @{}}
    \centering Lorie Liebrock \\
    \hrulefill \\ \centering Academic and Research Advisor \\
    \end{tabular}

    \vspace{0.5cm}
    
    \begin{tabular}[b]{@{} p{5.5in} @{}}
    \centering Dongwan Shin \\ 
    \hrulefill \\
    \end{tabular}\qquad

    \vspace{0.5cm}

    \begin{tabular}[b]{@{} p{5.5in} @{}}
    \centering Stewart Thompson \\ 
    \hrulefill \\
    \end{tabular}\qquad

    \vfill
    \centering I release this document to the New Mexico Institute of Mining and Technology.
    
    \vspace{0.5cm}

    \normalsize
    Emily Kesler \hfill April 2, 2024
    \normalsize
    \noindent \begin{tabular}{@{}p{5.5in}@{}}
    \hrulefill \\
    \end{tabular}

\end{center}

\newpage

\clearpage
\pagenumbering{arabic} 
\section{INTRODUCTION}

\indent

In 2023, the global average data breach cost per country was 4.45 million USD, according to a report from the International Business Machines Corporation (IBM) \cite{IBM-Stat}. Another company estimated that the total cost of cybercrime for the world in 2024 will be 9.5 trillion USD, which includes estimated fines, lost business, and remediation costs \cite{Cyber-Cost, Juniper-CS}. These shocking costs are not sustainable. The trends of cyber attacks and cybercrime are not in our favor. They grow increasingly more prominent every year. The field of cybersecurity is unique and ever-evolving. New threats, exploits, tactics, defenses, problems, and solutions are discovered daily. Due to the nature of the discipline, defenders almost always play catch-up to attackers, often only able to address issues that have already been used to exploit information systems. This poses many difficult challenges, severely affecting companies' and individuals' digital and, in most cases, financial security. Despite these challenges, cybersecurity experts press on, learning and developing new tactics, ideas, and solutions to protect critical information. This paper proposes a framework to assist cybersecurity experts with developing those solutions.

To be discussed, cybersecurity must first be defined. The U.S. Cybersecurity and Infrastructure Security Agency (CISA) published a blog post in 2021 where it defines `cybersecurity' as ``the art of protecting networks, devices, and data from unauthorized access or criminal use and the practice of ensuring confidentiality, integrity, and availability of information” — a simple, easy-to-understand definition \cite{CISA-def}. So, according to CISA, cybersecurity is an art dedicated to protection. That viewpoint is maintained throughout this paper. Furthermore, a few years before that blog post, Craigen, Diakun-Thibault, and Purse \cite{Craigen-et-al-2014} bemoaned that definitions of the term `cybersecurity' were, and are ``highly variable, context-bound, often subjective, and, at times, uninformative.” They point out that part of the difficulty comes from the nature of modern cybersecurity. That is, they believe modern cybersecurity is transdisciplinary, so the definition should reflect that. Despite the challenges, they created a definition that states: ``Cybersecurity is the organization and collection of resources, processes, and structures used to protect cyberspace and cyberspace-enabled systems from occurrences that misalign \textit{de jure} [perceived] from \textit{de facto} [actual] property rights” \cite{Craigen-et-al-2014}.

The proposed definition is intended to include the influence of other disciplines without stating it explicitly. Craigen, Diakun-Thibault, and Purse \cite{Craigen-et-al-2014} assert that the line “[...]the organization and collection of resources, processes, and structures[...]” is broad enough to capture the complex, overlapping disciplines involved in modern cybersecurity. This definition is suitable for ‘cybersecurity.’ It can also be used to define ‘transdisciplinary cybersecurity.’ Craigen, Diakun-Thibault, and Purse would argue that they are the same. Indeed, their definition is intended to include and define (and redefine) cybersecurity as something encompassing transdisciplinary cybersecurity. For the purposes of this paper, it is believed that additional context would assist in understanding the term. One of the references included by Craigen, Diakun-Thibault, and Purse is from Frederick Chang, the former Director of Research at the National Security Agency. Chang (2012) describes cybersecurity as being primarily about “adversarial engagement,” which, at its lowest level, is one person (or a group of people) defending something or someone from another person (or group of people). However, it is Chang’s final sentence in that reference that truly defines this paper's idea of transdisciplinary cybersecurity. ```In addition to the critical traditional fields of computer science, electrical engineering, and mathematics, perspectives from other fields are needed’” \cite{Craigen-et-al-2014}. \textit{Perspectives from other fields are needed.} At its core, that is the transdisciplinary approach to cybersecurity. Incorporating not just the expertise of brilliant engineers but also the expertise of brilliant psychologists, ethicists, or any other expert who might assist in bettering the art of protection that is cybersecurity.

Cybersecurity is traditionally associated with a rigid science discipline filled with computer scientists, mathematicians, and engineers. While it is true that those core areas and skills are essential to cybersecurity, modern cybersecurity has evolved beyond the boundaries of tradition. Technology’s exceptionally rapid proliferation across the contemporary world now means that every sector, both public and private, will have cybersecurity concerns. Cybersecurity is no longer a distinctly isolated discipline reserved only for technically trained or technology-focused individuals. It is made better by introducing the ideas and influences of other disciplines. 

One might wonder, why make it so complicated? Technology is technology, and cybersecurity is no different. There is no need to muddy the waters with other disciplines. There is credence to that point. The distinctive features of a discipline are precisely what makes it a discipline all its own. However, due to humankind's messy reality, technology is now more than just technology. It might have once been reserved for things like the U.S. government’s ARPANET project, but no longer. Technology today is a business owner’s billing software, ensuring they keep their bills paid and a roof over their head. It is a tool for designing spacecraft, permitting incredibly detailed, exact measurements. It is a life support system in a hospital, allowing someone to fight back against their illness for just a little longer. To better protect these precious pieces of technology and human life, the traditional idea of cybersecurity needs to be elevated. This paper challenges the assumption that cybersecurity is an exclusive, homogenous discipline. Cybersecurity, \textit{good} cybersecurity, is transdisciplinary.

Most cybersecurity literature focuses on presenting specific technical solutions to specific problems that cybersecurity experts and digital defenders face. Few papers provide guidance on applying a transdisciplinary method to cybersecurity. That literature gap is the purpose of this paper. This paper offers a framework and a starting point for the transdisciplinary cybersecurity process in a broader context. Taking inspiration from the business ethics-based stakeholder theory, the framework is intended to help experts shift into a transdisciplinary mindset when tackling problems. Bringing together multiple disciplines is challenging, but this framework will attempt to ease some difficulties by outlining steps and practices to encourage collaboration. In addition to stakeholder theory, the framework was developed using three guiding resources to create a good framework, which are discussed below in Section 2.1. Those resources’ principles are applied at every step of the development process for this framework. The framework is intended to provide a suitable foundation for future research and enhanced cybersecurity approaches and practices. However, the power of the framework ultimately lies in the execution of the ideas outlined in it. Following the framework in this paper, three use cases are presented to test the framework's effectiveness in real-world situations.

\section{BACKGROUND AND RELATED WORKS}

\indent

A review of cybersecurity literature revealed that the concept of transdisciplinary cybersecurity makes up a tiny portion of that literature. As mentioned, most cybersecurity literature presents specific solutions to specific problems. Some papers emphasize the need to include other disciplines \cite{Craigen-et-al-2014, Christen-et-al, Jacob-et-al, Schaeffer-2017, Zintgraff-2014, Ramirez-Choucri-2016, McNeese-2017, Brazell-2014}, but only a few, and not always in the same context. Of that discovered subset, most papers fall into one of two categories: academics, specifically how to develop a transdisciplinary cybersecurity curriculum for college students, or cyber-physical systems. 

The phrase `transdisciplinary' appears to be used interchangeably in the literature with `interdisciplinary' and `multidisciplinary.' This paper includes references that use `transdisciplinary,' `interdisciplinary,' and `multidisciplinary.' For the purposes of this paper, because so many authors use them interchangeably, they will refer to the same idea. However, it is necessary to note the difference between the terms. 

\begin{figure}[t]
    \centering
    \includegraphics[width=10cm]{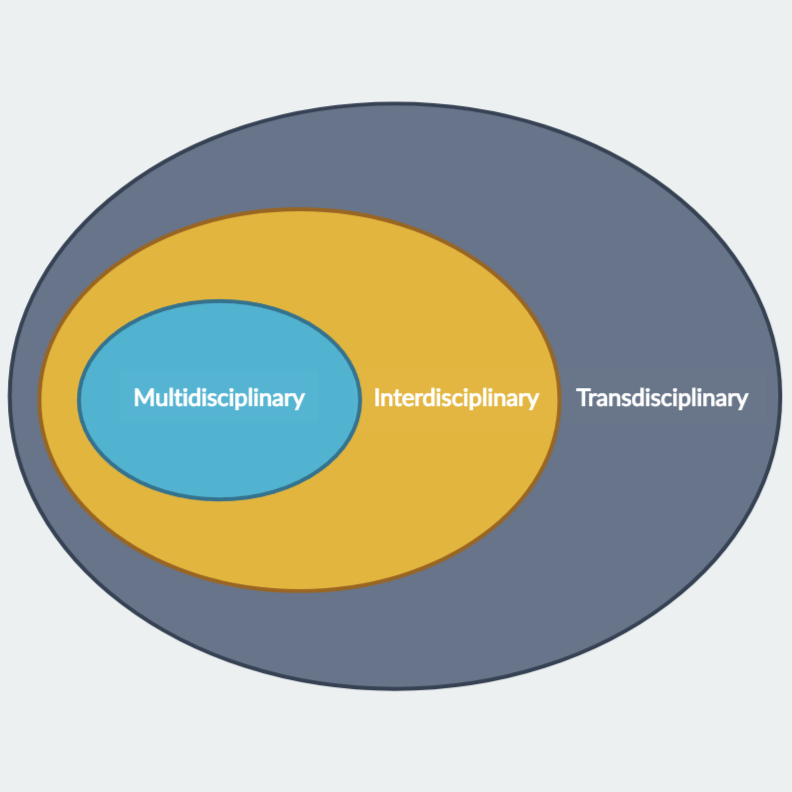}
    \caption{Multidisciplinary, Interdisciplinary, and Transdisciplinary Diagram}
    \label{fig:encap-diag}
\end{figure}

Utrecht University in the Netherlands \cite{Utrecht-NL} and Oklahoma State University’s (OSU) Research Toolkit for Librarians \cite{OSU-Toolkit} offer definitions of the three ideas for better understanding. According to OSU, \textbf{multidisciplinarity} ``draws on several disciplines in parallel but they remain separate from each other” \cite{OSU-Toolkit}. Utrecht University adds that it is simply studying the perspectives of multiple disciplines at the same time to gain a broader understanding of a subject \cite{Utrecht-NL}. \textbf{Interdisciplinarity} goes a step further than multidisciplinarity by integrating ``perspectives or insights from different perspectives [\textit{sic}] through interaction [...] to better understand a complex phenomenon” \cite{Utrecht-NL}. OSU states that interdisciplinarity ``synthesizes approaches from different disciplines into a new and coherent whole” \cite{OSU-Toolkit}. Finally, a \textbf{transdisciplinary }approach is ``about bringing together knowledge from science and practice” \cite{Utrecht-NL} and ``integrates and transcends disciplinary boundaries, bridging humanities and sciences” \cite{OSU-Toolkit}. This paper would like to add that transdisciplinarity brings together knowledge from other disciplines and transcends disciplinary boundaries but also enables an entirely new environment where creative solutions and methods may be discovered. So, while they are closely related and indeed build off of one another, these three terms are, in fact, separate approaches. Hence, the title of this paper aims for a transdisciplinary approach rather than a multidisciplinary or interdisciplinary approach. \textbf{Figure \ref{fig:encap-diag}} illustrates the relative scope of multi-, inter-, and transdisciplinary approaches (created in \textit{Creately}, Melbourne, Australia). Some of the works related to these concepts are discussed below.

Craigen, Diakun-Thibault, and Purse \cite{Craigen-et-al-2014} attempt to create a more inclusive, concise definition of ‘cybersecurity.’ They do this by establishing criteria for what their definition should demonstrate, chiefly: 1) other definitions of cybersecurity can be ``mapped into” their definition, and 2) their definition is ``unifying and inclusive in that it supports interdisciplinarity” \cite{Craigen-et-al-2014}. The authors perform a literature review, analyzing past definitions of cybersecurity to get to a complete definition that meets that criterion. They then extract dominant themes and distinguishing aspects from the selected literature and craft a new definition. Throughout the piece, the authors emphasize the transdisciplinary nature of cybersecurity and how a complete definition of the term should include that nature.

Focusing on the value-conflicts of cybersecurity, Christen et al. \cite{Christen-et-al} describe the hazards of modern cybersecurity decisions. They argue that more than the classic dichotomy of security versus privacy is needed for the contemporary ethical challenges practitioners of cybersecurity face. To support that argument, Christen et al. offer cybersecurity value conflicts in four areas: business, health, national security, and design processes. The paper also touches on the point that cybersecurity affects technology in all social domains and maintains that view throughout. Overall, the piece is a thoughtful discussion of cybersecurity decisions' far-reaching moral and ethical effects.

In their work, Jacob, Peters, and Yang \cite{Jacob-et-al} discuss the need for a cybersecurity curriculum to include more than just computer science or engineering-based scholars. From their perspective, the rapid growth of cybersecurity warrants a change to the present collegiate curriculum, specifically including other disciplines in that education. In presenting this idea, the authors discuss what other fields have contributed, and will contribute, to cybersecurity. They also point out that understanding the human element of cybersecurity is essential to training new professionals. This approach aims to train the next generation of cybersecurity experts who understand technical and non-technical skills and thus make better decisions based on that understanding. For further reading, in addition to Jacob, Peters, and Yang, several other authors were discovered to advocate for curriculum changes and incorporate other disciplines in secondary education \cite{Schaeffer-2017}, \cite{Zintgraff-2014}.

Ramirez and Choucri \cite{Ramirez-Choucri-2016} perform an extensive literature review of existing cybersecurity research from 2012 to 2015 and offer suggestions on cybersecurity terminology. One intent of the paper was to ``explore avenues of cyber security [sic] that have not received as much traditional attention as standard topics” \cite{Ramirez-Choucri-2016}. They assert that to better cooperate across disciplines, there needs to be more clearly defined standards for cybersecurity research. They believe that a more easily accessible and common vocabulary for use between experts is critical to that standardization. To help bridge the gap between disciplines, they offer some guidelines, terms, and metrics. The overarching theme of the piece is that cybersecurity requires collaboration. According to the authors, that collaboration begins with formalizing the terminology standards. Thus, the authors present their recommendations on how they believe one can do that better. 

With a focus on the human factor of cybersecurity, McNeese and Hall \cite{McNeese-2017} offer a framework that incorporates social aspects into cybersecurity, which they call socio-cyber systems. They point out that although technology is the focus of cybersecurity operations, human intelligence is what initiates and drives those operations. Due to that fact, McNeese and Hall emphasize the need to understand the cognitive and social factors that play into cybersecurity. To support their points, the authors performed a qualitative and quantitative study to understand an individual's mental processes in cybersecurity. One of the main goals of the framework they discuss is focused on increasing cyber situational awareness and resilience in the face of challenges. The authors believe their approach can contribute to positive outcomes such as enhanced asset protection, teamwork, understanding of events, and situational awareness.

Brazell \cite{Brazell-2014} discusses the need for transdisciplinary security of cyber-physical systems. Cyber-physical systems directly impact the physical world, such as heart pacemakers, vehicle anti-lock brakes, or industrial controls \cite{Brazell-2014}. According to Brazell, cyber-physical systems are ``at the heart of the design, manufacture, installation, operation, and sustainability of private corporate and national infrastructure across the world” \cite{Brazell-2014}. Not only are those systems essential, but they are also incredibly vulnerable to attacks and manipulation. He argues that those critical cyber-physical systems result from transdisciplinary engineering, so a transdisciplinary approach to security is only logical.

Jacob, Peters, and Yang \cite{Jacob-et-al} highlight three additional disciplines specifically that would aid in creating and maintaining a holistic perspective. They state that including the disciplines of criminology, legal studies or law, and economics would significantly enhance the learning, study, and practice of the field of cybersecurity. Christen et al. \cite{Christen-et-al} echo the sentiment and highlight technology, legal, philosophical, and social sciences as essential disciplines to consider. The authors of this paper would like to include political science, psychology, computer science, engineering, physics, and chemistry as additional areas of interest. The reasoning for the inclusion of these areas is discussed below:
\begin{itemize}
    \item Criminology, psychology, philosophy, and law further enhance the understanding of the human element of cybersecurity.
    \item Political science and economics are included due to the far-reaching effects of cybersecurity incidents, from the impact of data breaches on U.S. Social Security funds to the regulatory bodies and standards for cybersecurity and technology to the accessibility of technology for all income levels.
    \item Engineering is the broad term for the collection of engineering specialties such as electrical, chemical, or mechanical engineering. Many engineering specialties naturally border cybersecurity due to the design and understanding of components, function, and maintenance. 
    \item Computer science is a discipline that naturally borders cybersecurity.
    \item While the natural sciences of chemistry and physics may seem out of place, they can provide experts with a greater understanding of situations, other viewpoints, and potential solutions. For example, in the case of chemistry in electric vehicles, one could exploit the volatility of the battery through a cyber vulnerability that would jeopardize the vehicle's passengers or any bystanders near the vehicle.
\end{itemize}

\begin{figure}[t]
    \centering
    \includegraphics[width=10cm]{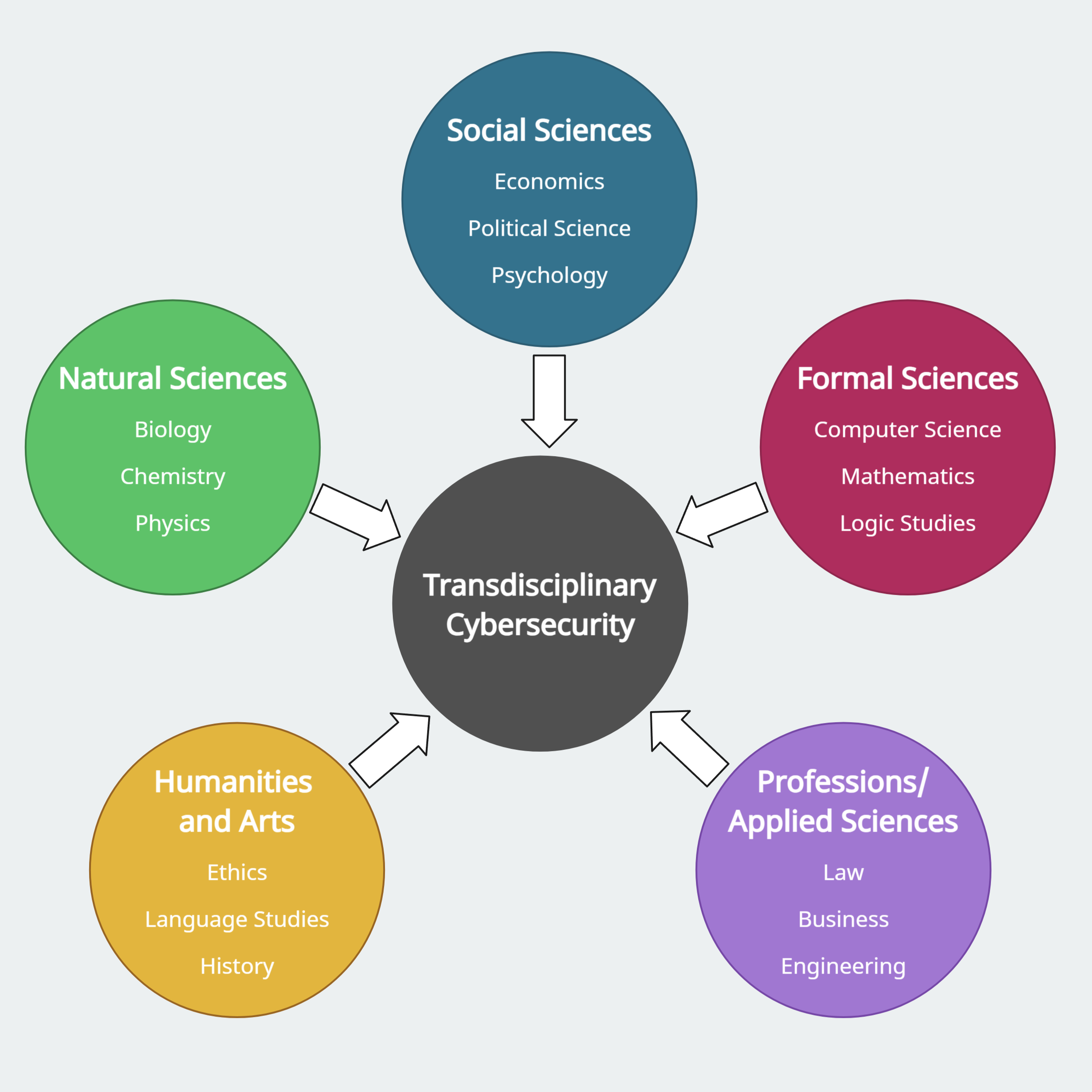}
    \caption{Transdisciplinary Influences Diagram}
    \label{fig:mindmap}
\end{figure}

A visualization of disciplines was researched and created as part of this analysis. This diagram is intended to help readers understand transdisciplinary cybersecurity. The visualization draws influence from the topics named by Jacob, Peters, and Yang, Christen et al., and a mind map of academic disciplines from the GoGeometry website by Antonio Gutierrez \cite{Gutierrez}. Gutierrez’s mind map contains five branches of academic disciplines: social sciences, natural sciences, formal sciences, professions/applied sciences, and humanities and arts. Those five branches, in addition to several of the sciences under each of those classifications, are pictured in \textbf{Figure \ref{fig:mindmap}}.

    \subsection{What Makes A Framework}
    
    \indent
    
    To create a framework, one needs to understand a framework. In researching this subject, the World Wide Web turned up a plethora of websites, webpages, and other sources, all offering opinions about how to create a framework for their specific discipline. The uses for the frameworks range from software to business, but most of the frameworks this search discovered were meant for software developers. As this is not a paper focused on software development, those resources are not directly relevant to the type of framework that this paper wishes to develop. 

    Apart from the resources available, three specific sources were selected to guide the development of this project \cite{KU, OECD-Framework, Wallace}. These were chosen due to their instruction and approach, which pair well with the idea of a transdisciplinary cybersecurity framework. Each resource offered framework definitions alongside their steps and guidance. Additionally, University of Kansas \cite{KU} and Wallace \cite{Wallace} provided examples of the frameworks that aligned with the traits and instructions outlined in their resources. The last source is a framework designed for public governance \cite{OECD-Framework}, which offered a functional, complete reference on which to base this paper’s framework. By its very nature, this paper’s transdisciplinary framework means combining the ideas of those different sources to create something usable.
    
    The University of Kansas (KU) \cite{KU} hosts a website titled ``Community Tool Box” that contains a variety of teaching, training, and guidance for individuals or groups hoping to organize and develop their community. One of the ``Toolkits” in this Tool Box is instruction on developing a framework (or model of change). While more focused on social sciences, this toolkit presents a logical approach to outline, appropriately scope, and develop a fully fleshed-out framework. Key to this toolkit is the idea of a logic model, which can be produced using another piece of the KU Community Tool Box \cite{KU-logic}. KU defines a logic model as something that ``presents a picture of how your effort or initiative is supposed to work” \cite{KU-logic}. Leveraging the use of logic models, the KU framework development consists of nine individual steps to assist organizers in creating a viable framework. Each of the main steps is listed below, but \cite{KU} provides additional insight on their website with several subsections and questions below each main point.
    \begin{enumerate}
        \item Describe the intended uses of your framework or model of change.
        \item Outline your initiative or program's vision and mission.
        \item State the objectives of your initiative or effort.
        \item Describe the appropriate scope or level of your framework or model of change.
        \item Identify ALL components to include in the logic model or model of change.
        \item Using the components, draft a picture of the framework or model of change.
        \item Check for the completeness of your logic model.
        \item Once all current components and elements are identified and incorporated into the framework or logic model, put it to use.
        \item Revise the model (as needed) to adapt the elements and incorporate newly emerging ones.
    \end{enumerate}
    
    The development of the transdisciplinary framework began with the guidance provided by KU's Community Toolbox. The application of the steps and process used are further discussed in the Methodology section, Section 4, of this paper.

    The Organization for Economic Cooperation and Development (OECD) is an international organization devoted to creating and maintaining better public policies worldwide. The OECD publishes numerous guides, results of their studies, and other documents to support governments in making better policies across all sectors. One such publication, the \textit{Policy Framework on Sound Public Governance}, has been instrumental in developing this project and its framework \cite{OECD-Framework}. While the OECD framework is for policy formulation, implementation, and evaluation, it provides an excellent example of a functional framework. The formatting, presentation, and language used in the guide were incredibly useful in shaping, evaluating, and updating the transdisciplinary approach framework featured in this paper. The OECD framework also advocates for cultivating a values-based culture for public policy. This focus on values, such as integrity, inclusiveness, and accountability, aligns well with the concepts and motivations for this project.
    
    Wallace \cite{Wallace} discusses the general requirements for any framework, not one specifically suited to any particular field or discipline. He defines a framework as ``a tool for organizing information.” He then includes six characteristics of a good framework:
    \begin{itemize}
        \item It has a specific use.
        \item It helps to identify the necessary information and it filters out the unnecessary information.
        \item It is simple to use.
        \item It provides usable insight.
        \item It encourages and rewards deeper learning.
        \item It encourages more effective communication.
    \end{itemize}
    This paper seeks to emulate the six characteristics listed above, and the transdisciplinary framework iterations were checked consistently against said characteristics to ensure quality.
    
\section{METHODOLOGY}

\indent
    
Formulation of the framework began with the steps provided on KU’s Community Tool Box page \cite{KU}. It should be noted that the first part of this process used steps 1 to 5 of the KU guidance. Steps 6 through 9 were used later in the design process once the logic model was further developed. Upon reaching the end of step 5 of the KU process, the other two guiding resources, \cite{OECD-Framework} and \cite{Wallace}, were incorporated with the ideas generated by the KU approach. This was done to ensure that all three resources were included in the brainstorming sessions before a formal model of the framework was finalized. 
    
    \begin{description}
        \item[Step 1.] The framework's purpose, direction, and intended use were identified. The purpose of the framework is to \textit{encourage cybersecurity experts to work with and learn from other disciplines to improve their cybersecurity approaches}. This purpose pairs with the direction to \textit{promote and achieve transdisciplinary approaches by creating an adaptable framework grounded in cybersecurity principles}. The intended use of the framework is threefold. The uses are listed in Section 4 but are also included here. The framework is designed to be used \textit{as an evaluation tool for existing cybersecurity practices or postures}, \textit{as a development tool to engage with other disciplines to foster learning and create new solutions}, and \textit{as a guidance tool to encourage new ways of thinking about, perceiving, and executing cybersecurity practices}.
        \item[Step 2.] In this step, the framework’s vision and mission were summarized. The vision for this framework is \textit{a tool that brings together a community of experts devoted to working together to better protect everyone under their care}. The mission is to \textit{support the development of more robust and applicable cybersecurity postures by creating meaningful conversations between field experts and encouraging the inclusion of multiple perspectives for enhanced learning and application}. This step was essential in defining a single target goal for the framework. That is, \textit{to create a tool that brings together a community of experts devoted to working together, protecting those under their care, and enabling them to make better cybersecurity decisions}.
        \item[Step 3.] This step walked through how to define the framework's objectives. The objectives for this framework include \textit{enhanced cybersecurity teaching, learning, implementation, community collaboration, and new, creative solutions to complex problems and cyber threats inspired by discussions between field experts}. With the ideas from Step 2 and the objectives listed, assumptions and factors that contribute to the problems and goals of this concept were considered. Identifying the specific, measurable results of the framework was especially helpful to idea generation. The objectives further refined the vision and ultimate purpose of the framework: \textit{to create a community of experts devoted to working together, who can participate in meaningful conversations with other discipline experts, and enable everyone to make positive, well-grounded cybersecurity decisions}.
        \item[Step 4.] The desired scope of the framework was selected from one of the three options. While both the overall initiative (option A. of Step 4) and selecting a particular initiative (option B. of Step 4) were conducive to brainstorming, the last level mentioned, a specific work plan for action (option C.), is the most applicable to this framework. It was determined that the framework cannot be overly broad in scope for effectiveness. However, it was also determined that the framework needs to be flexible enough for many potential uses. Trying to balance both the size and flexibility of the framework in this phase was difficult. Lastly, at this step, the first concepts of individual framework pieces were devised to be refined later.
        \item[Step 5.] This was one of the more difficult steps. This is because the guide asks one to outline all components that should be included in the model of change or, in this case, the framework. The items listed in the guide are the intended framework's purpose, context, inputs, activities, outputs, and effects. The framework’s purpose, or mission, was modified and restated: \textit{to support the development of more robust and applicable cybersecurity postures by creating meaningful conversations between discipline experts, encouraging the inclusion of multiple perspectives for enhanced learning and application, and enabling everyone to make positive, well-grounded cybersecurity decisions.} Following that, the context and conditions for the framework’s creation were summarized as \textit{there is very little literature on transdisciplinary cybersecurity currently}, and \textit{there are widely varying fields, experiences, and situations for each cybersecurity expert}. Potential inputs identified include \textit{a pre-existing cybersecurity plan or posture for evaluation} and \textit{specific goals or focus areas that cybersecurity experts want to strengthen}. The activity, or intervention, is the framework itself. The expected outputs, based on the potential inputs above, are \textit{an updated, well-rounded cybersecurity posture or approach} and \textit{new approaches or solutions for the focus areas or goals selected}. Finally, the effects were categorized as short-term, mid-term, and long-term outcomes. Those expected outcomes include \textit{an up-to-date assessment of an organization’s cybersecurity posture or practices}, \textit{an improved understanding of cybersecurity and the transdisciplinary approach}, and \textit{the development of relationships with other field experts for future projects}.
    \end{description}

    With the groundwork laid, the development of the framework structure began. The characteristics recommended by Wallace heavily influenced this stage, primarily to make the framework simple to use and a framework that encourages and rewards deeper learning \cite{Wallace}. The desire for simplicity birthed the concept of three single-word stages (Think, Plan, Do) that could guide and encourage framework users without overwhelming them. Following the division of the framework into three pieces, the individual steps for each stage were developed. It was not easy trying to find a balance between creating something simple and insightful but not so simple or broad that it became unhelpful. Drawing inspiration from the Transdisciplinary Cybersecurity curriculum at New Mexico Tech, which offered new perspectives on disciplines (such as the ethics of cybersecurity), the steps for each stage were created. The first step in the \textbf{Think} stage was easy to devise, as were the two \textbf{Do} items. However, working on the remaining three steps in \textbf{Think} and the three steps in \textbf{Plan} was more complicated. Again, drawing from Wallace \cite{Wallace}, it was decided that no more than four steps should occupy each stage of the Think, Plan, Do approach so as not to overcrowd the framework and make it difficult to use. The missing steps were created by asking: How does one lead people to think about other disciplines? Inspired by Morgan and Gordijn’s stakeholder approach \cite{Morgan-Gordijn-2020}, the next logical step after identifying a problem was to consider stakeholders, which are vital to helping experts understand the complete picture. Then, following that, the other steps were devised to lead framework users towards thinking in a transdisciplinary manner. Once each stage was completed, the initial framework skeleton was reviewed, and the wording was modified to make each step more straightforward and adhere to Wallace’s guidance. Then, the OECD framework was used as an example, in conjunction with Wallace’s characteristics, to populate each stage and its accompanying steps. The OECD framework was particularly crucial to devising and formatting the framework’s intended uses in this paper \cite{OECD-Framework}. At this point, a complete first draft of the framework was presented to advisors so they could offer feedback, and the framework could be modified and polished for its final version.

    \subsection{Motivations}

    \indent
    
    Readers have likely already discerned that this paper and framework are heavily value-driven. The values have defined this work's intent, approach, and presentation. Due to that fact, discussing the motivations has been deemed essential. Three fundamental values are the central drivers behind this project and paper. 
    \begin{enumerate}
        \item Every human being deserves respect until their actions prove the contrary. By extension, their ideas, perspectives, and talents deserve respect.  
        \item The purpose of cybersecurity is to protect, whether it be people, physical objects, or digital objects.
        \item People can achieve more when they work together. There are beautiful cases where a single person has achieved spectacular things. But more often than not, the great people of history and modern day have been assisted, influenced, or mentored by at least one other person (such as a parent or parental figure).
    \end{enumerate}
    
    \subsection{Across Disciplines}

    \indent
    
    Though it is stating the obvious, people are not perfect. Despite these imperfections, humanity has managed to reach this point in history. There is a veritable library of possible reasons as to why that might be. One reason is very simple but profound: humans learned how to work together for a common cause. Despite differences of opinion, political leanings, social status, and other separating labels, people have learned to combine their efforts to achieve great things. Is it not then reasonable to state that cooperation is paramount to the survival of the human race? Cooperation, collaboration, and camaraderie between disciplines and their concepts will strengthen cybersecurity. Is it easy to listen to another’s ideas or perspectives? Not always. Will it be complex, messy, and incredibly frustrating to work with and implement the views of other disciplines into cybersecurity? Most definitely. But when has a first attempt at an invention or technique ever been without challenge? Never. So, too, will it be with this attempted approach to transdisciplinary cybersecurity. But tenacity and cooperation of people in the face of adversity is what has allowed humanity to flourish. Tenacity and cooperation are what will enhance cybersecurity beyond heights not yet seen.
    
    \subsection{Stakeholder Theory}

    \indent

    One of the largest inspirations for this project is the concept of stakeholder theory. Specifically, the inspiration comes from a paper by Morgan and Gordijn \cite{Morgan-Gordijn-2020}, which applies stakeholder theory to cybersecurity ethics. Stakeholder theory is an approach to business ethics introduced by Edward Freeman in 1984 \cite{Stanford-ethics}. This theory was developed in direct opposition to the shareholder theory, which dictated that businesses should seek only to maximize their shareholders’ wealth \cite{Stanford-ethics}. In contrast, the earliest versions of stakeholder theory asserted that instead of working only for the profit of a business’ shareholders, managers should balance the interests of all stakeholders involved, where a stakeholder is defined as anyone who has a “stake” or interest (financial or otherwise) in the business \cite{Stanford-ethics}. To summarize, shareholder theory argues that business revenues are to be used to maximize shareholder wealth. In contrast, stakeholder theory argues that the revenues are to be used to benefit all stakeholders, not just shareholders. The Stanford Encyclopedia of Philosophy's entry on ``Business Ethics” asserts that both stakeholder and shareholder theory are not as crucial as the moral constraints involved in business decisions \cite{Stanford-ethics}. That means that a corporate manager should balance the interests of both shareholders and stakeholders to achieve an acceptable outcome. 
    
    Though detractors would argue that stakeholder theory is not so much a theory as it is a collection of ideas, the concept has been refined significantly since its creation \cite{Stanford-ethics}. That is not to say the theory is without flaws, but the refined approach is much more applicable and practical than the original concept proposed in 1984. That refined stakeholder theory is utilized by Morgan and Gordijn, who apply a “care-based stakeholder approach” to cybersecurity in a business setting \cite{Morgan-Gordijn-2020}. In an evaluation of the theory’s evolution, Morgan and Gordijn share that the original developers of stakeholder theory believe there is a need to replace conflict with communication, cooperation, and collective action \cite{Morgan-Gordijn-2020}. This paper is not supposed to be a moral authority; it does, however, attempt to adapt and apply the reasoning behind this improved stakeholder theory to cybersecurity.

    As Morgan and Gordijn step through the “care-based stakeholder approach,” they describe three fundamental questions that are presented and answered by the original developer of the approach, Daniel Engster \cite{Morgan-Gordijn-2020}. These questions are based on and around business ethics; however, the underlying principles of the corresponding answers apply to this paper. Each question posed by Engster (and presented by Morgan and Gordijn) has been given its own section below for easier understanding.

    \noindent \textbf{``Who exactly counts as a stakeholder?”}
    
    This first question establishes a necessary foundation of definition. The answer defines a stakeholder as one whose functioning and survival are tied directly to the business and its activities \cite{Morgan-Gordijn-2020}. Those individuals are specifically listed, ``namely, shareholders, employees, the local community, customers, suppliers, and competitors” \cite{Morgan-Gordijn-2020}. That is a much narrower scope of stakeholders than the original stakeholder theory, which would want any individual affected by the business to be counted as a stakeholder. It is (practically) noted by the authors that using the original theory’s definition of a stakeholder makes meeting stakeholder needs impossible: a business would exhaust all resources trying to meet those needs, thus preventing care of those most closely engaged with it \cite{Morgan-Gordijn-2020}. 

    \noindent \textbf{``How should businesses distribute care to those stakeholders?”}
    
    Three ethical principles are offered in response to that question: the proximity principle, the relational principle, and the urgency principle \cite{Morgan-Gordijn-2020}. The proximity principle dictates using limited resources to first care for individuals who are close before caring for those that are farther away \cite{Morgan-Gordijn-2020}. This principle can be likened to emergency instructions on a commercial aircraft. If there is a loss of cabin pressure, oxygen masks will drop, and passengers must first secure their own masks before assisting others. Morgan and Gordijn also state that it can be argued that the proximity principle justifies caring for oneself first, caring for individuals who are geographically or temporally closer to oneself, and caring for individuals in one’s own culture or state before those in foreign cultures or states \cite{Morgan-Gordijn-2020}. The second principle, the relational principle, says businesses should prioritize caring for individuals they have close personal relationships with over others \cite{Morgan-Gordijn-2020}. Engster defines a close relationship as a relationship where one party depends on another to meet their survival or developmental needs, such as a parent and child \cite{Morgan-Gordijn-2020}. Though business relationships are not always as directly linked as a parent and child might be, Engster argues that these close relationships ``deserve priority because they are so closely tied with the goals of caring” \cite{Morgan-Gordijn-2020}. Lastly, the urgency principle states that businesses should care for individuals with more urgent needs before seeing those with less urgent needs \cite{Morgan-Gordijn-2020}. Urgency is determined by considering what effect an action or inaction could have on a person or group’s survival \cite{Morgan-Gordijn-2020}. Engster mentions that ``if there is a focus on the urgent needs of stakeholders over less urgent ones, this allows a business to prioritize the needs of individuals or groups who will not survive or function without acting” \cite{Morgan-Gordijn-2020}. Morgan and Gordijn conclude that the urgency principle also reduces the number of stakeholders to be considered, making the distribution of care, time, and resources more feasible for a business \cite{Morgan-Gordijn-2020}.

    \noindent \textbf{``What ethical approaches should businesses adopt when conflict arises amongst stakeholders?”}
    
    The simple answer to this question is that the highest priority should be given to shareholders and employees because, according to Engster, their interests are generally more important than other stakeholders’ interests \cite{Morgan-Gordijn-2020}. However, Morgan and Gordijn mention that this does not necessarily apply in all situations. Engster includes an overriding condition in his paper: the health and safety of employees and customers \cite{Morgan-Gordijn-2020}. If the health and safety of either of those groups is jeopardized, their interests take precedence. Engster continues and states that keeping individuals employed should be a favored goal of a business, at least in the short term \cite{Morgan-Gordijn-2020}. There is, of course, a need for balance so as not to destroy a business at the cost of not cutting any jobs. But if a cut is unavoidable, businesses are urged to use the ‘rule of consensus’ where a company tries to find a solution to stakeholder conflicts ``that are acceptable to all by communicating the proposed solutions to stakeholders and trying to solicit alternative proposals from them” \cite{Morgan-Gordijn-2020}.

    To summarize the information, a stakeholder is defined, three ethical principles are offered to determine which stakeholders receive resources (and how much), and guidance is provided to help navigate the inevitable conflicts between different stakeholders. Some of the information presented might seem like common sense to readers, but the explanation of these different ideas by Morgan and Gordijn helped nurture the concept of this project. Those ideas are evident in the underlying skeleton of the framework and present in some of the methods and steps in the final product, particularly in the approach created and used in Section 4.
    
\section{THE FRAMEWORK}

    \indent

    While the framework’s primary target audience is cybersecurity practitioners, the concepts within may resonate with individuals outside the discipline. The framework aims to lay a foundation for and encourage a community of collaboration, continual learning, and transdisciplinary thinking.
    
    In application terms, the framework is expected to be used:
    
    \begin{itemize}
        \item As an evaluation tool for existing cybersecurity practices or postures.
        \item As a development tool to engage with other disciplines to foster learning and create new methods.
        \item As a guidance tool to encourage new ways of thinking about, perceiving, and executing cybersecurity practices.
    \end{itemize}
    
    This framework is not intended to be a penultimate instruction manual on transdisciplinary thinking in cybersecurity. There is no perfect transdisciplinary approach. It is understood that what works for one individual or team may not work for others. However, this framework is designed to be a stepping stone toward developing a professional culture where transdisciplinary thinking is the norm in cybersecurity and other fields.
    
\begin{figure}[t]
        \centering
        \includegraphics[width=10cm]{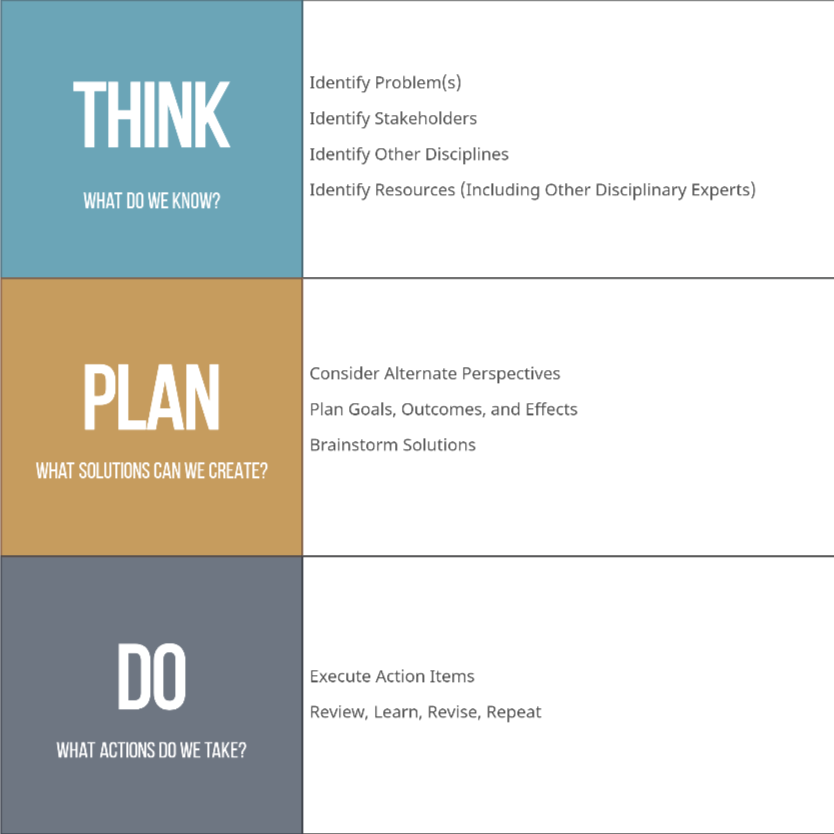}
        \caption{The Think, Plan, Do Approach}
        \label{fig:tpd-approach}
    \end{figure}

    The framework uses a three-step approach, aptly named the Think, Plan, Do approach, visualized in \textbf{Figure \ref{fig:tpd-approach}}. While simple, this approach is intended to help facilitate transdisciplinary thinking and the development of new solution methods.

    Before beginning the framework discussion, it is worth including a short explanation of the realist approach presented by Louart et al. \cite{Louart-et-al-2023}. They discuss an approach to policy evaluation, but what they offer can be applied to understanding this framework. In their work, they remark that a policy causes effects, but not directly \cite{Louart-et-al-2023}. Policies may or may not trigger mechanisms, which may or may not trigger outcomes \cite{Louart-et-al-2023}. It is in the analysis of these mechanisms, or ``the ways people react to the resources, sanctions or opportunities (depending on the type of intervention) made available to them,” and their influence on the outcomes that constitute the realist approach to evaluation \cite{Louart-et-al-2023}. Following that thought, this framework is expected to trigger mechanisms leading to outcomes. Louart et al. comment that people are the drivers of change, so it will be their action, inaction, or reaction to the ideas proposed and generated in this framework that will produce outcomes, whether they be positive or negative, expected or unexpected \cite{Louart-et-al-2023}. 

    \subsection{THINK}

    \indent
    
    The first section of this framework is \textbf{Think}. This first portion of the framework will guide individuals to identify the problem(s), identify stakeholders, and identify potential resources or other disciplines they can call on to learn from. The central question for this section is determining \textit{``What do we know?”}

        \subsubsection{Identify the Problem(s)}

        \indent
        
        The first step of \textbf{Think} is to identify the problem, or alternatively the point, of the effort or issue being tackled. Is the framework being used to evaluate an existing cybersecurity posture or practice? Is it being used to develop new methods and engage with other disciplines on a particular topic? Is it being used to try to create new perspectives, solutions, or solution methods for something? What is the ultimate goal of this effort? Determining the problem, or point, of the effort and the ultimate end goal will offer individuals a solid foundation for devising solutions for their efforts. Finally, perhaps most importantly, what are the needs and goals of the organization utilizing the framework? 

        \subsubsection{Identify Stakeholders}

        \indent
        
        Next, stakeholders must be identified. Who will be directly affected by this effort? Who would benefit from it? Who would suffer from it? If the list of stakeholders seems to be growing too large, Morgan and Gordijn’s definition of a stakeholder will help keep the scope of the effort within its bounds \cite{Morgan-Gordijn-2020}. It would also be prudent to consider the following: What potential conflicts may arise between stakeholders, and how might one navigate those conflicts?

        \subsubsection{Identify Other Disciplines}

        \indent
        
        With the stakeholders identified, it is time to consider other disciplines that could offer insight into the effort. Does the effort naturally border other fields (such as cybersecurity for power systems or healthcare)? If so, what are they? Were potential attackers identified as stakeholders in the effort? A psychologist or human behavioral expert consultation could contribute additional understanding. Does the effort touch on a person’s rights (such as a cybersecurity posture that requires individuals to forego their privacy)? An ethicist or legal professional may help clarify questions surrounding the effort. It may be helpful for team members to review the Transdisciplinary Influences Diagram in \textbf{Figure \ref{fig:mindmap}} for additional ideas. Brainstorming efforts may generate unrelated possibilities, but this step aims to cast the net wide so it might be refined later.

        \subsubsection{Identify Resources (Including Other Disciplinary Experts)}

        \indent
        
        At this point, the next step is to identify the resources available to the effort. Is the effort reliant on a budget, and if so, how much is available? If the effort is timebound, how much time does it have for completion? What equipment is provided or needed for the effort? Identify key team members. Who must be involved in the effort in order for it to succeed? Then, consider who might be a valuable resource inside or outside the organization. If no one is known, who can be contacted to connect team members to those potential resources? Are there any other tools or methods available that might expedite or assist the effort? In that same vein, are there any collaboration tools available? These tools can serve to connect individuals and streamline the collection of ideas.
        
    \subsection{PLAN}

    \indent
    
    Following the assessment in the \textbf{Think} phase, the next stage of the approach is \textbf{Plan}. This step may be similar to the \textbf{Think} phase in that it encourages brainstorming, but the focus differs. Users are encouraged to consider alternate perspectives in their planning, develop goals, outcomes, and desired effects, and then brainstorm and plan the solution(s) to reach those goals. The central question of this section is, \textit{``What solutions can we create?”}

        \subsubsection{Consider Alternate Perspectives}

        \indent
        
        With stakeholders, other disciplines, and resources identified, it is time to consider alternate perspectives of the problem or effort. 

        The first step in this process is acknowledging the limits of one’s personal or group perspective. It is crucial to accurately assess one’s biases, experience, and areas of expertise. Only with acknowledgment and humility can an individual or group place themselves in the right state of mind to fully consider alternate perspectives. That is not to say that individuals who struggle with seeing alternate views are arrogant or not trying. It is a learned skill that will develop over time as individuals work to integrate alternate perspectives and ways of thinking into their own.

        With a solid understanding of an individual or group's limitations, gaps in perspective become more apparent. Individuals should work to spot those gaps, whether in perspective or knowledge.
    
        Once gaps are found, individuals can utilize other disciplinary experts or examine resources (identified in the Think phase or discovered in this step) to glean insights that can be added to the final solution. Questions might include: What does [the resource] think of this problem? How do they approach it? Is that approach viable in this context? How does it differ from this team’s perspectives? How might that concept be incorporated into a solution?
 
        \subsubsection{Consider Goals, Plan Outcomes, and Discuss Effects}

        \indent
        
        At this point, individuals are advised to revisit the overall goal of the effort. Recentering and refocusing on the effort's ultimate goal(s) will allow teams to further refine their ideas for potential approaches and solutions that align with the final desired outcome. Individuals or groups may have discovered additional goals that they would like to include in the effort and, as such, can add those goals to their considerations.

        With the effort's goals revisited and refined, individuals can plan the desired outcomes that match the ultimate goal(s). Using the realist approach from Louart et al. \cite{Louart-et-al-2023}, the mechanisms and context of a given effort are essential to understanding and accurately planning the outcomes of that effort. As a refresher, the realist approach argues that policy creates effects or outcomes, but not directly \cite{Louart-et-al-2023}. A policy (or, in this case, the effort) triggers mechanisms that then lead to outcomes \cite{Louart-et-al-2023}. So, the mechanism is a vehicle for the policy to reach the desired outcome. Giving attention to the intermediate mechanisms in planning can offer insight into the processes and actions that will be used to achieve the effort’s ultimate end. Questions for identifying the intermediate mechanisms can include: What is the context surrounding this effort? How are people expected to react to the effort or the implementation of the effort? Will those reactions positively or negatively impact the effort, and how so?

        Analysis of the intermediate mechanisms pairs with the discussion of the effort's potential effects (intended or unintended). That discussion can offer additional information for better understanding and planning considerations. Contributions from the alternative perspectives identified above can help individuals further contemplate the potential effects of the effort.

        \subsubsection{Brainstorm Solutions}

        \indent
        
        With the resources and considerations gathered in the \textbf{Think} stage and the concepts already discussed from the \textbf{Plan} phase, individuals can brainstorm potential solutions and, most importantly, how to reach them. Solutions may have already been discussed, deliberated, or proposed. However, at this step, individuals can now focus entirely on devising solutions to the problem(s) they defined at the beginning of the framework. With resources, concepts from other disciplines, and the intended outcomes, goals, and effects defined, individuals have the tools to revise or create solutions or solution methods specific to their needs.  
    
    \subsection{DO}

    \indent
    
    The final stage in the Think, Plan, Do approach is to \textbf{Do}. In this stage, individuals assign specific action items, execute the actions, review the effort, learn from the review results, revise the approach, and repeat as necessary. The overarching question for this final step is \textit{``What actions do we take?”}

        \subsubsection{Execute Action Items}

        \indent
        
        With a solution planned, action items should be developed and assigned. The work done in the Think and Plan phases will amount to nothing if no action is taken. Activities should be coordinated between the relevant stakeholders and then assigned and executed. Cooperation, collaboration, and camaraderie should underpin the effort and actions taken. Individuals can and should leverage the resources available to effectively apply and achieve the goal(s) of the effort. To restate the realist approach again, actions taken lead to mechanisms, which, in turn, lead to outcomes. Utilizing, understanding, and adjusting to the mechanisms between action and outcome will assist individuals in effectively achieving their ultimate effort goal(s).

        \subsubsection{Review, Learn, Revise, Repeat}

        \indent
        
        Four steps contribute to a successful effort execution: review, learn, revise, and repeat. As action items are checked off and the effort progresses, it is essential to review the effort. The three guiding resources \cite{KU, OECD-Framework, Wallace} used for this framework touch on the concept and importance of review and revision. With accurate data from a review, individuals can identify weak spots and learn from the information presented. After identifying potential improvement items, revisions can be made as necessary. Then, finally, repeat those first three steps. A cyclical review, learning, and revision process will offer insight into the effectiveness of the effort and foster a culture of continual learning. 

\section{USE CASE EXAMPLES}

\indent

Three use cases of real-world examples have been developed to showcase how the framework might be used. Each use case pertains to one of the expected uses of the framework: as an evaluation tool, as a development tool, and as a guidance tool.
     
To reiterate, the framework is intended as a tool to generate new ways of thinking, new approaches, and new solution methods to both ordinary and unique problems. Its use in these situations is limited only by the author's limited experience. In the hands of experts, it is hoped that this framework can generate even more creative and innovative methods and solutions. The \textbf{scenarios are imaginary}, but the \textbf{referenced cyber attacks are events that have occurred in the real world}. The references for those attacks are included in the respective scenarios. Each use case defines a situation and then applies the framework to that scenario.

    \subsection{Evaluation Tool Example: Caesars Entertainment}

    \indent
    
    The following scenario is for the framework’s intended use as \textit{an evaluation tool for existing cybersecurity practices or postures}.
    
    \textbf{Scenario:} The Chief Information Security Officer (CISO) of Caesars Entertainment, a large hotel and casino management company, has been tasked with evaluating the organization's current information security policy and updating all business processes to match any policy changes. The request comes following a successful ransomware attack against the organization, severely impacting the internal information technology (IT) infrastructure and casino operations. Findings from the ransomware attack indicate that a social engineering attack on an outsourced IT support vendor’s phone allowed attackers to access and lock down company systems \cite{CPO-Mag, CYBS-Dive}. The ransomware gang, the Scattered Spider threat group, initially demanded payment of \$30 million from Caesars Entertainment \cite{CPO-Mag}. The ransom was negotiated to \$15 million, and Caesars Entertainment paid that reduced price to unlock their systems \cite{CPO-Mag}. Still, Caesars Entertainment wants to avoid another costly payout, so the CISO has been tasked with evaluating the policy to see what might be improved.
    
        \subsubsection{THINK: Identify Problem(s)}

        \indent
        
        The problem or point of the evaluation is the current information security policy. The framework will be used as an evaluation tool for the policy. The assessment is needed because of a recent breach of the organization’s information security that cost valuable time, money, and customers. Caesars Entertainment needs to prevent something like that from happening again, or at the very least, minimize the negative fallout should it occur in the future. 

        \subsubsection{THINK: Identify Stakeholders}

        \indent
        
        The stakeholders directly affected by this effort are the organization’s shareholders, employees, customers, their families, competitors, the suppliers and third-party organizations supporting the organization, and the ransomware attacker(s). Everyone belonging to the organization is believed to benefit from this evaluation and the improvement of company information security. The ransomware attackers, or any other outside actors, will suffer for this improvement. Competitors to the organization may benefit or suffer from the policy updates based on their perspective of and intent towards Caesars Entertainment.

        \subsubsection{THINK: Identify Other Disciplines}

        \indent
        
        Drawing from the mindmap of disciplines \cite{Gutierrez}, several fields of study and the reasoning behind their selection were identified. The disciplines involved in this effort include: 
        \begin{itemize}
            \item Information Technology (IT) – For the affected IT systems.
            \item Computer Science – For the obvious breach in cybersecurity.
            \item Psychology – This is for several items: the perception of the company by customers and competitors, the attacker's logic and reasoning, and the third-party vendor’s thought processes and reasoning.
            \item Sociology – This is possibly also for the perception of the company and the attacker’s logic/reasoning, as well as the relationships between the company, competitors, customers, vendors, and attackers.
            \item Legal or Law Studies – For the laws broken by the attackers and the legal damage control the organization may have to do because of the breach.
            \item Business – For the organization’s business management of all facets of this ransomware problem, both internally and externally, as well as the Business Continuity Plan.
            \item Journalism and Mass Communications – For the public relations and press management that the organization must do or has done regarding the attack.
            \item Ethics - As the policy applies to people, it is vital to ensure that it is ethical (according to one's desired ethical perspective).
        \end{itemize}

        \subsubsection{THINK: Identify Resources (Including Other Disciplinary Experts)}

        \indent
        
        Caesars Entertainment management has stated they are willing to expand the information security budget by an additional \$600,000 for any improvements that need to be made. The policy evaluation must be completed as soon as possible to prevent a duplicate attack from either the same or another ransomware gang on the already weakened infrastructure. The CISO has a team of dedicated information security experts who are ready and willing to implement any controls that are deemed necessary for additional infrastructure protection. Additionally, the CISO reaches out to the Chief Information Officer (CIO) of Caesars Entertainment to gain insight into the Information Technology (IT) side of the Business Continuity Plan and Disaster Recovery Plan. The CISO contacts an outside policy expert to assist in evaluating the policy. That policy expert then provides access to a policy lawyer and a psychologist who can offer additional insights into understanding the policy process and attacker/defender reasoning.

        \subsubsection{PLAN: Consider Alternate Perspectives}

        \indent
        
        With the resources they’ve identified, the CISO begins evaluating the policy. Working with the policy expert, the CISO finds some problems with the existing policy.
        \begin{enumerate}
            \item The language surrounding contractor and third-party security is vague, and the decisions for security are ultimately left up to the individual vendor.
            \item The policy requires new employees to participate in security training before accessing organizational systems, and all employees must annually participate in security training, but third parties do not need to participate in the training before touching the organization’s systems.
            \item There are no defined intervention strategies for management to address a potential insider threat from an employee or vendor.
            \item The information security policy directs readers to the Business Continuity Plan and Disaster Recovery Plan for action during and recovery from serious security incidents; neither plan includes processes or recovery options in the event of a ransomware attack.
            \item The current infrastructure does not require multi-factor authentication (MFA) for users or administrators to access the organization's systems.
            \item The Information Security team is only required to perform minimal system log monitoring, with no mandated consistent schedule for log reviewing.
        \end{enumerate}
        
        With these glaring issues identified, the CISO and policy expert open dialogues with the CIO, policy lawyer, and psychologist for further input. 
        
        The CIO and CISO work together to identify points in the IT infrastructure that can be improved. From the CIO’s perspective, security improvements cannot be allowed to overburden the existing infrastructure. Considering that, the CISO works with the CIO to brainstorm potential agreeable solutions for both parties. The CISO helps the CIO update the Business Continuity Plan and Disaster Recovery Plan to include processes for ransomware attacks, and the CISO decides to have more information about both plans in the information security policy.
        
        The policy lawyer emphasizes the need for the policy to be ``airtight” to prevent lawsuits in the event of another external or internal breach. That means updating the language in the policy and implementing new controls to cover the gaps identified by the CISO and policy expert. The policy lawyer also points the CISO to resources to help them update the language.

        The psychologist is able to talk through the ransomware gang’s reasoning and willingness to commit crimes to gain something, whether it be fame or fortune. With their input, the CISO can understand and think about how to make the organization less appealing to potential outside attackers. The psychologist also recommends some behavioral providers and strategies for dealing with a person of concern to help mitigate an insider threat.

        \subsubsection{PLAN: Plan Goals, Outcomes, and Effects}

        \indent
        
        Revisiting it, the effort’s goal is to create an updated security policy that prevents or mitigates the potential fallout from a ransomware attack. Ideally, the revised security policy will address the identified gaps and help improve the organization’s infrastructure security. Regarding effects, there is likely to be pushback from some managers and third-party organizations regarding the policy requirements for third-party security. It may also be costly and time-consuming to push out security training to all third-party contractors interacting with the organization’s systems.

        \subsubsection{PLAN: Brainstorm Solutions}

        \indent
        
        To address the identified issues, the CISO developed potential solutions with input from the CIO, policy expert, policy lawyer, and psychologist. They first created specific solutions for each item identified in the policy review.
        \begin{enumerate}
            \item Implement new security controls, namely the principle of least privilege, to restrict and remove unnecessary third-party vendor access to company systems.
            \item Train all third-party vendors given access to organization information systems on information security and social engineering basics.
            \item Work with human resources (HR) to train management in intervention strategies to minimize the risk of an insider threat.
            \item Coordinate with the CIO to designate a responsible party to update and maintain the Business Continuity and Disaster Recovery Plans, ensuring they include appropriate actions for cyber attacks or other cyber events.
            \item  Research potential MFA solutions, find a suitable solution, incorporate MFA requirements into the policy, and implement MFA to access organizational systems.
            \item Update the requirements for system log monitoring, particularly the robustness, standardization, and frequency of system log reviews.
        \end{enumerate}
        
        In addition to those six solutions for each of the previously identified policy gaps, the CISO and their experts brainstormed additional ideas that may also be beneficial.
        \begin{itemize}
            \item Modify the policy's language, with guidance from the policy expert and organizational lawyers, to create a policy that can help protect the organization in the face of potential lawsuits.
            \item Update the annual security training for employees to include social engineering and how they might avoid or stop it.
            \item Research and, if within budget, implement network micro-segmentation to minimize the threat of an attacker being able to pivot through the organization’s internal systems.
            \item Research and identify ways to make the organization less appealing to potential attackers.
            \item Research and bring in (or find one in the organization) a business ethicist to ensure the policy meets ethical standards.
            \item Hire a change management consultant to assist with the organizational transitions caused by the security policy. 
            \item Perform a cost-benefit analysis (CBA) to determine the best options for MFA and the other suggested security controls.
        \end{itemize}

        \subsubsection{DO: Execute Action Items}

        \indent
        
        From the list of potential solutions, the CISO has identified which ones are the most viable, namely: 
        \begin{itemize}
            \item Implement new security controls, namely the principle of least privilege, to restrict and remove unnecessary third-party vendor access to company systems.
            \item Work with human resources (HR) to train management in intervention strategies to minimize the risk of an insider threat.
            \item Coordinate with the CIO to designate a responsible party to update and maintain the Business Continuity and Disaster Recovery Plans, ensuring they include appropriate actions for cyber attacks or other cyber events.
            \item Research potential MFA solutions, find a suitable solution, incorporate MFA requirements into the policy, and implement MFA to access organizational systems.
            \item Update the requirements for system log monitoring, particularly the robustness, standardization, and frequency of system log reviews.
            \item Modify the policy's language, with guidance from the policy expert and organizational lawyers, to create a policy that can help protect the organization in the face of potential lawsuits.
            \item Update the annual security training for employees to include social engineering and how they might avoid or stop it.
            \item Research and identify ways to make the organization less appealing to potential attackers.
            \item Research and bring in (or find one in the organization) a business ethicist to ensure the policy meets ethical standards.
            \item Hire a change management consultant to assist with the organizational transitions caused by the security policy. 
            \item Perform a cost-benefit analysis (CBA) to determine the best options for MFA and the other suggested security controls.
        \end{itemize}

        \subsubsection{DO: Review, Learn, Revise, Repeat}

        \indent
        
        Following the execution of the above actions, the CISO revisits the outcomes to see if anything needs modification.
        \begin{itemize}
            \item The new security controls have been implemented, but one of the requirements to remove third-party vendor access has made much more work for the information security team. That responsibility either needs to be delegated or automated (say, the deactivation of an account after 14 days of inactivity).
            \item HR has successfully trained management on some intervention strategies. A few managers have already used them on their employees.
            \item With support from the CISO, the CIO has designated a Business Continuity Plan Coordinator and Disaster Recovery Plan Coordinator to review and update the plans as necessary.
            \item Upon performing a CBA, the results helped the organization select a feasible MFA solution. They have begun rolling out MFA to all its IT systems. The rollouts are being performed in small batches, but so far, the infrastructure has been able to support the changes, with employees being trained ahead of the rollouts.
            \item With the updated requirements for log monitoring, the Information Security team has modified the robustness of the gathered logs and has a set schedule to review system logs. The team has also begun researching improved log monitoring tools and solutions they might implement.
            \item The policy's language has been updated and is much clearer now, protecting the organization from potential lawsuits regarding ransomware attacks.
            \item The annual security training has been updated, and phishing drills have been implemented to keep employees on their toes and constantly aware of potential social engineering attacks.
            \item The organization is working with an outside cybersecurity firm to find ways to make themselves less appealing to attackers.
            \item The organization consults an ethical expert to review the security policy to ensure it meets general ethical standards. The ethicist suggests several word changes, specifically regarding privacy. With input from the organization’s legal counsel, the ethicist approves the policy once the changes are deliberated, modified, and executed.
            \item The change management consultant was tasked with developing and implementing change strategies and initiatives. The transition for the organization has had a few hiccups, but overall, it progressed smoothly, thanks to the consultant’s help.
            \item An annual policy review has been scheduled so updates or modifications can be done as needed.
        \end{itemize}
        
    \subsection{Development Tool Example: Electric Vehicle Cybersecurity}

    \indent
    
    The following scenario was developed to showcase the framework used as \textit{a development tool to engage with other disciplines to foster learning and create new methods}.
    
    \textbf{Scenario:} The United States Department of Transportation’s (USDOT) National Electric Vehicle Infrastructure (NEVI) Formula Program is a federal funding initiative that aims to support the development of a national electric vehicle (EV) charging infrastructure across all interstate corridors \cite{USDOT-NEVI}. The New Mexico Department of Transportation (NMDOT) has successfully submitted and received funding for the state’s development of electric vehicle infrastructure. However, following the 2015 Ukrainian power grid attacks, some power companies are reluctant to work with NMDOT to develop something that could jeopardize their critical infrastructure \cite{CISA-Ukr}. To help soothe these fears of cyber attacks, NMDOT has hired a private IT security firm to assist them with the project. A Team Lead has been assigned to develop a plan to bring the power companies on board with creating the EV infrastructure.

        \subsubsection{THINK: Identify Problem(s)}

        \indent
        
        This effort aims to encourage power companies to collaborate with NMDOT to plan electric vehicle usage throughout the state. However, power companies fear the security risks associated with the power grid/critical infrastructure and the unknowns of electric vehicles, their systems, and their integration with the power infrastructure. The framework will be used as a development tool to help the IT security firm and the NMDOT strike up conversations between stakeholders and other disciplines to confront this challenging issue.

        \subsubsection{THINK: Identify Stakeholders}

        \indent
        
        Stakeholders directly impacted by this effort include the NMDOT, the power companies, the private IT security firm, EV charging station vendors, consumers who use EVs, the U.S. Joint Office of Energy and Transportation responsible for planning nationwide EV infrastructure, members of the U.S. DOT Federal Highway Administration (FHWA) who oversee the NEVI program funding, and the Western Electric Coordinating Council (WECC), which oversees the North American Electric Reliability Corporation Critical Infrastructure Protection (NERC CIP) program for public utilities in the western United States on behalf of the Federal Energy Regulatory Commission (FERC). Due to the size of the agencies, the IT security firm immediately determines the need to enlist a professional change management consultant to facilitate a workshop for the stakeholders to discuss the group’s actionable goals. It is believed that the inclusion of all these organizations will be beneficial. However, because NERC’s mission is ``to assure the effective and efficient reduction of risks to the reliability and security of the grid,” \cite{NERC} the private IT security firm believes the inclusion of representatives from the WECC under NERC will be especially helpful in assuaging the fears of the power companies who are hesitant.

        \subsubsection{THINK: Identify Other Disciplines}

        \indent
        
        The Team Lead from the private IT security firm reviews the list of stakeholders and identifies the disciplines involved in this effort. Those disciplines include:
        \begin{itemize}
            \item Economics – Power companies and the NMDOT must not only determine the costs for consumers associated with the EV infrastructure, but this effort will also impact the state and national economies regarding EVs, their sales, power provisions, public infrastructure, and insurance costs.
            \item Political Science – Because the NEVI program is a federal initiative, policies dictate the awarding and use of federal funding. Additionally, public policy experts must evaluate the public’s acceptance of and demand for EV infrastructure.
            \item Law, specifically Regulations – Many regulations govern the EV infrastructure space, such as those on energy, cybersecurity, or critical infrastructure.
            \item Ethics – As this is a program for public infrastructure, the ethics issues involved, such as ensuring fair implementation in underserved socioeconomic communities and maintaining EV owner privacy across the entire EV ecosystem, need to be assessed and addressed.
            \item Business – This effort involves public and private interests, specifically the private interests of EV charging businesses. Business managers across disciplines must also determine the best course of action for the program to benefit all stakeholders while maintaining safety and security for their customers and organizations. Another business consideration is the competing interests between electric vehicles and the gas and oil industries. 
            \item Computer Science – The cybersecurity efforts of all stakeholders who cooperate in the EV initiative are crucial for protecting customers, critical infrastructure, and businesses. This may also include Operational Technology (OT) and Internet of Things (IoT) security.
            \item Information Technology – This applies to the information systems used to administer services in conjunction with the OT devices that monitor and deliver power to those systems.
            \item Electrical Engineering – This field’s inclusion is an obvious choice due to the nature of EVs.
            \item Control Systems Engineering – The OT and Industrial Control Systems (ICS) that monitor and provide power to consumers are vital to this effort.
            \item Physics – This effort can and should involve scientific research into the delivery methods for vehicles and engineering.
            \item Chemistry – As with the inclusion of physics research, this effort will be reliant upon scientific research in chemistry for the safe operation of EVs throughout the planned public infrastructure. Specifically, this means studying and researching lithium batteries or other power sources to not endanger individuals who buy, operate, and are near EVs.
        \end{itemize}

        \subsubsection{THINK: Identify Resources}

        \indent
        
        Upon the Team Lead’s recommendation for a workshop between stakeholders, the NMDOT selects a trusted change management consultant to coordinate and facilitate the workshop. Due to the size of the effort, a planning committee is necessary to determine available resources from each agency for the workshop. Working with the consultant, the Team Lead creates a planning committee with individuals from each stakeholder agency. In addition to identifying the resources available to their respective organizations, the stakeholders must also consider:
        
        \begin{itemize}
            \item Budget – What is the total budget for the workshop?
            \item Timeline – What is the timeline and the expected completion date for a final recommendation after the workshop, and who will be drafting the recommendation?
            \item Equipment – What equipment is necessary for the workshop to be successful?
            \item Operations – Will additional staff be required? Will additional software be needed for the effort? How much will that cost?
            \item Logistics – Who is responsible for providing the venue, agenda, and staff for the effort?
        \end{itemize}

        Over the course of two separate virtual planning meetings, the committee coordinates and confirms all resources they can devote to this effort. Upon completing the planning meetings, the stakeholders are comfortable with the management consultant being the logistical and operational manager for the workshop. The Team Lead from the private IT security firm is responsible for developing relevant and appropriate content for the workshop.

        \subsubsection{PLAN: Consider Alternate Perspectives}

        \indent
        
        Armed with a list of stakeholders, the groups who will participate in the workshop, and ample resources, the Team Lead works to consider alternate perspectives.

        The NMDOT wants to move forward with this project and bring EVs to New Mexico. They want to develop and create a safe, sustainable public infrastructure that enables EV ownership both in the state and across the nation. The NMDOT is depending on the private IT security firm to bring the hesitant power companies on board so that the project can continue.

        The local power companies’ primary focus is maintaining the availability of the power they provide to their clients through reliability. They are concerned that EV charging stations, systems, and power consumption will exacerbate current or create new risks to the existing critical infrastructure. The power companies do not trust the third-party EV charging vendors enough to allow them to attach their charging systems to the power infrastructure.

        The private IT security firm wants to adequately address the concerns of the power companies and assist in facilitating a successful workshop. The project Team Lead wants to create a discussion between all stakeholders that will benefit their client, the NMDOT, and ensure the safety and security of the critical infrastructure. The Team Lead intends to gather a list of cybersecurity considerations specific to power companies supporting EV infrastructure, relevant standards, and cybersecurity compromises in similar cyber ecosystems worldwide to develop an agenda for the workshop.

        The change management consultant is focused on planning and facilitating an impactful, successful workshop. Based on the responses of the involved parties, the consultant seeks to facilitate future meetings for NMDOT as well.

        The U.S. Joint Office of Energy and Transportation’s ultimate goal is to ``support the deployment of zero-emission, convenient, accessible, equitable transportation infrastructure” \cite{DOE-DOT}. They seek to support the meetings between local governments and their power providers to contribute to a nationwide infrastructure. In addition, the Joint Office can leverage its laboratory partners, Pacific Northwest National Laboratory (PNNL) and Idaho National Laboratory (INL), to contribute to a productive cybersecurity discussion.

        The representatives from the U.S. DOT FHWA are first and foremost focused on the goals of the NEVI program. The program aims to develop and deploy a nationwide EV charging infrastructure along major corridors and create an interconnected charging station network that will allow for improved data collection, access, and reliability \cite{USDOT-NEVI}. In addition to those goals, the representatives will seek to ensure the NMDOT’s compliance with the standards necessary for NEVI funding.

        The WECC is an organization devoted to maintaining the security and reliability of the bulk power system (BPS) in the Western Interconnection region under NERC \cite{About-WECC}. Under NERC, WECC has been given the authority to administer to power providers by monitoring compliance with and enforcing the standards set by NERC and FERC. The WECC representatives are focused on ensuring that the security and reliability of the BPS will not be compromised by the installation of EV supply equipment (EVSE) on the power grid.

        There are a number of different vendors who can and will provide EVSE for the NMDOT’s EV ecosystem. The primary goal of the vendors is to generate revenue for their respective businesses. They will seek to find the most cost-effective methods of developing and installing their systems. They want the power companies to come on board so they (the vendors) can move forward with providing supplies for this ambitious EV infrastructure project.

        Consumer concerns about EV infrastructure vary widely. Some support the project, and others are against it for a variety of social, political, and personal reasons. Primarily, though, consumers who own or want to own an EV are focused on their personal safety, the safety and cost of their vehicles, and preventing the theft of their credit card information.

        \subsubsection{PLAN: Plan Goals, Outcomes, and Effects}

        \indent
        
        Another virtual meeting between stakeholders is held to determine the workshop's goals, outcomes, and planned effects. After deliberation, all attendees agree on the following list:
        \begin{enumerate}
            \item Identify and develop a comprehensive list of all existing relevant standards involved in the power and EV infrastructure effort. This list must be reviewed and compared against any planned implementations.
            \item A design reference architecture must be developed and then agreed upon.
            \item A table of known cybersecurity attack vectors and their recommended mitigations must be developed and shared with decision-makers.
            \item A framework for cybersecurity policies and procedures, such as the National Institute of Standards and Technology Special Publication (NIST SP) 800-82, must be selected and used consistently across stakeholder efforts.
            \item A list of actions that relevant parties can take to mitigate stakeholders' security and cybersecurity concerns must be developed and integrated into the stakeholders' final recommendation report at the conclusion of the workshop.
            \item Improved trust and communication must be developed between local power companies and the NMDOT.
            \item If the workshop is determined to be fruitful, another workshop session should be scheduled for more stakeholder discussions.
        \end{enumerate}

        \subsubsection{PLAN: Brainstorm Solutions}

        \indent
        
        A workshop is held with the support of the agencies, the change management consultant, and the private IT security firm. Over three days, the workshop discusses the first five planned items from the list above.
        \begin{enumerate}
            \item Utilizing their knowledge and resources, the workshop develops a list of relevant standards for their effort. Some of the applicable standards include:
                \begin{itemize}
                    \item \textbf{The National Electric Vehicle Infrastructure Standards and Requirements} – A rule from the FHWA on NEVI minimum requirements.
                    \item \textbf{Society of Automotive Engineers (SAE) J1772} – A standard on EV plugs, couplers, and outlets that defines Type 1 charging connectors (commonly referred to as J Plugs).
                    \item \textbf{International Electrotechnical Commission (IEC) 61850-90-8} – The 61850 standard defines communications for substations, and 90-8 defines communications for e-mobility in EVSE.
                    \item \textbf{IEC 62196} – Standards for EV plugs, couplers, and outlets, which defines Type 1 and Type 2 charging connectors.
                    \item \textbf{IEC 68151} – A standard for electric vehicle charging systems.
                    \item \textbf{International Organization for Standardization (ISO) 15118} – An international standard for road vehicles and vehicle-to-grid (V2G) communications.
                    \item \textbf{Institute of Electrical and Electronics Engineers (IEEE) 802.11 and 802.15.4} – Standards for cellular, Zigbee, and wireless communications.
                    \item \textbf{IEEE 1547} – A standard that defines interconnections for Distributed Energy Resources (DERs).
                    \item \textbf{IEEE 1901 and 1901.2} – Standards for communications over power lines.
                    \item \textbf{IEEE 2030.5, 2030.7, and 2030.8} – Standards that define communications between Distributed Energy Resources (DERs) and utilities, microgrids, and testing microgrids.
                    \item \textbf{NIST IR.8294} – A publication on cybersecurity research for EVSE.
                    \item \textbf{NIST IR.8473} – A Cybersecurity Framework Profile for EV Extreme Fast Charging (xFC) infrastructure.
                \end{itemize}
            \item Working with one another and with input from the assembled resources and standards, the workshop discusses a design reference architecture that they can present to the hesitant power companies.
            \item Leveraging the Joint Office’s connection to national laboratories, the IT security firm assists the workshop with drafting a table of known cybersecurity attack vectors and their mitigations. The laboratory partners offer past research on cybersecurity for EV infrastructure as a place for the stakeholders to start their cybersecurity planning \cite{Johnson-et-al}. The report includes its own list of known problems, recommendations, and mitigations relating to EV charging infrastructure.
            \item The workshop selects the International Society of Automation ISA/IEC 62443 standards for their cybersecurity policy framework. They believe the standards will be well suited to their needs, mainly because they include a mechanism for certifying products and vendor development processes, which will be of great importance for the EV infrastructure \cite{ISA-62443}.
            \item The workshop discusses potential actions that can be taken to address security and cybersecurity concerns. Some of the actions include:
                \begin{itemize}
                    \item Requesting or performing a risk assessment from the private IT security firm focused on any area of concern for stakeholders, such as a risk assessment of a vendor’s software or a power aggregation point.
                    \item Developing a standardized approval system for EVSE cybersecurity. The Joint Office will handle enforcement and may designate another agency to assist them.
                    \item Working with one or multiple national laboratories to develop and decide research projects addressing stakeholders' concerns.
            \item The workshop attendees designate the representatives from the IT security firm and the national lab partners as the responsible parties for developing and presenting a recommendation report that will be agreeable for all stakeholders.
                \end{itemize}
        \end{enumerate}

        \subsubsection{DO: Execute Action Items}

        \indent
        
        On the workshop's final day, the change management consultant leads the attendees in assigning roles and responsibilities for action items.
        
        \begin{itemize}
            \item The NMDOT, with input from the private IT security firm and national laboratories, will develop a sample reference architecture and present it to the WECC and Joint Office representatives before offering it to the power companies.
            \item The local power companies are responsible for identifying their top three major concerns and will provide them to the IT security firm for risk assessments and to the national laboratories for potential research on those areas of concern.
            \item The private IT security firm is responsible for performing risk assessments on the requested EVSE and EV infrastructure items. They, along with input from the national lab partners, will advise the power companies and vendors on cybersecurity best practices related to the EV infrastructure project. Together, the security firm and the national lab partners will create a report of recommendations that satisfies all stakeholders, which will be presented at the next workshop.
            \item The U.S. Joint Office will begin working on a standardized EVSE cybersecurity evaluation system and will update the stakeholders every six months.
            \item The FHWA will take the workshop materials and format and create a formalized system for collaboration on EV infrastructure between stakeholders, locally and across state lines.
            \item The WECC will report on each stakeholder’s goals and progress to the NERC, who will present it to the FERC for accountability of the larger federal agencies and offices.
        \end{itemize}
        
        At the end of the meetings, it was decided that the workshop had proved beneficial, and another workshop was scheduled to convene in six months. 
        
        \subsubsection{DO: Review, Learn, Revise, Repeat}

        \indent
        
        In six months, another developmental workshop was held. With efforts from the private IT security firm and national lab partners, the hesitant power companies were appeased enough that the development of new EV infrastructure could begin. 

        The U.S. Joint Office created mock-up standards for EVSE cybersecurity evaluations and sent the documents to the research community for comments. 

        The FHWA began asking states and local power companies to convene workshops so they could address their respective concerns and work towards creating a nationwide infrastructure for EV charging.

        The WECC collects quarterly reports from each workshop attendee. The reports are passed on to NERC and FERC to ensure the workshop's goals and action items are progressing as intended.
        
        The stakeholders in New Mexico will continue to collaborate in a three-day workshop each year, sharing their individual lessons learned and identifying areas for improvement.
            
    \subsection{Guidance Tool Example: National Laboratories Information Technology Summit}

    \indent
    
    The final scenario is intended to showcase the framework used as \textit{a guidance tool to encourage new ways of thinking about, perceiving, and executing cybersecurity practices}.

    \textbf{Scenario:} The National Laboratories Information Technology (NLIT) Summit is an event for U.S. Department of Energy (DOE) laboratory professionals that focuses on IT, cybersecurity, operations, technology, policies, and practices in support of the research performed at the national laboratories \cite{NLIT}. A cybersecurity expert has been selected to present at this conference. As part of their presentation, they want to provide a workshop to foster transdisciplinary thinking in cybersecurity research.

        \subsubsection{THINK: Identify Problem(s)}

        \indent

        The presenter begins the workshop by asking participants why they are attending this event. Some offer heartfelt answers, like seeking to learn more about cybersecurity or expanding their toolbox for addressing problems. Others offer humorous responses such as “My boss told me to come.” 

        The presenter then introduces the workshop's goal: to help participants think more broadly about their problems and assist them in devising comprehensive, impactful solutions. To do that, they will present a framework designed to encourage users to think more broadly and in a transdisciplinary manner. The participants are shown an image of \textbf{Figure \ref{fig:tpd-approach}} while the presenter briefly touches on the framework’s three stages and their accompanying steps. With a glimpse of the overarching framework, the presenter then asks participants to identify some common problems or challenges cybersecurity and IT specialists face. Answers ranged from general to specific. Some of the challenges include:
        \begin{enumerate}
            \item Hardware failure
            \item Software bugs or errors
            \item Keeping up with the most recent technology (staying on the cutting edge)
            \item Budget constraints
            \item People
            \item Compliance with laws and regulations
            \item Policies that are difficult to follow or implement
            \item Procedures that slow down operations or research
            \item Cybersecurity in general, but defending systems from attackers, both external and internal
        \end{enumerate}
        
        After identifying some problems, the presenter asks the participants to give short answers on how they might solve them.
        \begin{enumerate}
            \item Hardware failure – \textit{Maintain system backups, plan for redundancy in the IT infrastructure, and standardize the procurement process to get replacements quickly.}
            \item Software bugs or errors – \textit{Contact the developer. If the organization is the developer, identify the problem, code a solution, and push out the fix.}
            \item Keeping up with the most recent technology (staying on the cutting edge) – \textit{Developing and maintaining a forward-facing research and development division.}
            \item Budget constraints – \textit{Stay within budget, or at least do your best to stay within budget.}
            \item People – \textit{Employ consistent monitoring for security purposes, set up an IT support system to funnel calls or requests, or let the Human Resources (HR) department deal with them.}
            \item Compliance with laws and regulations – \textit{If the organization does not already have one, assign a Chief Compliance Officer (CCO) or other person to ensure compliance with all current regulations.}
            \item Policies that are difficult to follow or implement – \textit{Work with the policy writer(s) to modify, update, or remove the policy.}
            \item Procedures that slow down operations or research – \textit{Identify if the procedure is critical for operations, and if so, identify ways that it might be made less cumbersome and implement those changes.}
            \item Cybersecurity in general, but defending systems from attackers, both external and internal – \textit{Have sufficient security controls and procedures in place, consistently monitor systems for any changes or problems, stay in contact with the organization’s physical security providers, create and maintain a Continuity of Operations Plan (COOP), and keep an updated Disaster Recovery Plan.}
            \end{enumerate}
            
        The presenter selects three ideas for the workshop participants to focus on: hardware failure, people, and defending systems from attackers, externally and internally. The participants are then split up into three groups, each one assigned one of the problems above. They are then asked to drill down and devise a specific scenario within those three areas.
        \begin{enumerate}
            \item The first group chooses the hardware failure of a mission-critical server. The goal of their effort is to minimize the negative impact on the organization as a result of the failure.
            \item The second group identifies people who need to follow directions given by IT administrators (such as a procedure for passwords) but do not do what they are asked to do. This group aims to have people listen to instructions and follow proper procedures to improve the organization's security as a whole.
            \item The third group identifies an external attack from hackers seeking to breach company systems and steal customer information to sell on the black market. The goal of this group is to defend against the attack and protect their organization’s data.
        \end{enumerate}

        \subsubsection{THINK: Identify Stakeholders}

        \indent
        
        With the problems assigned, the presenter asks each group to identify stakeholders in their respective problem situations.
        
        For the first group (hardware failure), the following individuals were listed as stakeholders:
        \begin{itemize}
            \item IT staff who administrate the server
            \item The organizational operations affected by the mission-critical server
            \item Any staff who rely on the mission-critical server to get their job done
        \end{itemize}
        
        The second group (people) identified the problem’s stakeholders as:
        \begin{itemize}
            \item The person not following the directions or procedures
            \item The IT staff and administrators who need the person to do what they ask
        \end{itemize}
        
        The third group (defending systems from attackers) listed their stakeholders as:
        \begin{itemize}
            \item The organization being attacked
            \item The IT or cybersecurity staff who work as defenders
            \item The organization’s employees who might be affected by the attack
        \end{itemize}
        
        Each group presents the individuals they identified as their stakeholders. Then, the presenter gives the participants Morgan and Gordijn’s (2020) definition of a stakeholder: one whose functioning and survival are tied directly to the organization and its activities, ``namely, shareholders, employees, the local community, customers, suppliers, and competitors” \cite{Morgan-Gordijn-2020}. With the new definition, groups are asked to re-evaluate their lists of stakeholders to see if there are additional stakeholders they should include and whether the stakeholders would suffer or benefit from their problems' goals. With guidance from the presenter, who assists in identifying specific people or groups of people as stakeholders, each group revises its stakeholder lists.
        
        The first group (hardware failure) now has:
        \begin{itemize}
            \item IT staff who administrate the server who will benefit from the effort
            \item Organizational staff whose work is reliant on the mission-critical server who will benefit from the effort
            \item The organization’s shareholders who will benefit from the effort
            \item The organization’s customers will also benefit
            \item The suppliers to the organization will benefit from the effort
            \item The organization’s competitors may benefit from or suffer from the effort based on their perspective of the organization
        \end{itemize}
        
        The second group (people) has:
        \begin{itemize}
            \item The person not following the directions or procedures who will benefit from the effort, even if they don’t like it at first
            \item The IT staff and administrators who need the person to do what they ask, who will benefit from the effort
            \item The organization’s employees (who must all follow IT’s directions/procedures) who will benefit 
            \item The organization’s shareholders, who will benefit 
            \item The organization’s customers, who will also benefit from the effort
        \end{itemize}

        Finally, the third group (defending systems from attackers) identified:
        \begin{itemize}
            \item The organization’s shareholders, who will benefit from the effort
            \item The organization’s employees who will benefit as well
            \item The IT or cybersecurity staff who work as defenders will benefit
            \item The organization’s customers (whose data is being sought) will benefit from the effort 
            \item The organization’s suppliers (who are reliant on the business continuing to do business and protect the information it holds) will benefit
            \item The organization’s competitors may benefit or suffer from the effort
            \item The hackers who are attacking the information systems will suffer from the defense of the system if they are unsuccessful
        \end{itemize}

        \subsubsection{THINK: Identify Other Disciplines}

        \indent
        
        Following each group’s identification of their problem, the goal of their effort, and the stakeholders involved, the presenter provides each group with a copy of the mindmaps from Gutierrez and asks them to consider and identify disciplines that might be included in their respective efforts \cite{Gutierrez}. 
        
        The first group (hardware failure) chooses:
        \begin{itemize}
            \item Information Technology (IT) – For hardware failure and maintenance.
            \item Computer Science – Also for the hardware and maintenance.
            \item Computer Engineering – For the hardware implementation, maintenance, and server design.
            \end{itemize}
            
        The second group (people) selects:
        \begin{itemize}
            \item Psychology – For the individual(s) that the IT administrators have to talk to and direct.
            \item Sociology – Again, for the individuals that the IT team assists and directs.
            \item Information Technology (IT) – For the IT administrators and their procedures.
            \item Business – For the business or organization’s procedures that must be followed.
        \end{itemize}
        
        Lastly, the third group (defending systems from attackers) chooses:
        \begin{itemize}
            \item Computer Science – For the cybersecurity and defense of information systems.
            \item Business – For the impacts the attack would have on business or organizational operations.
            \item Psychology – For the attackers’ reasoning and motivations, perhaps even their attack patterns.
        \end{itemize}

        These choices are correct and very relevant, but they are limited. As the presenter walks through each group’s options, they encourage the participants to remember the identified stakeholders. They emphasize that the point of this exercise is to help them think outside the box and beyond the classical answers and solutions that all engineers or IT specialists think of. The presenter then adds additional disciplines to each group’s scenario that might be of interest.
        
        For the first group, the presenter suggested they consider the following in addition to what they previously identified:
        \begin{itemize}
            \item Business – Because the hardware failure will impact a mission-critical server, the business, its goals, its revenue, its budget, and its needs should be included in the discussion.
            \item Design – The design of the server, its connected systems, and the procedures surrounding that mission-critical asset are essential to ensuring the organization continues to function correctly.
            \item Communications – Specifically telecommunications and interpersonal communications because if a critical piece of the business goes down, people will need to know, especially to quickly resolve the situation.
            \item Law or Legal Studies – If this server is indeed mission-critical, there could be legal ramifications for the organization and the individual employees should the entire server fail, if it is not handled correctly, or if nothing is backed up.
            \item Ethics – If the mission-critical server experiences a hardware failure, it will impact the individuals reliant on that server. Those individuals deserve to be treated ethically and equitably in the event of a failure and should be notified of the failure and the progress of the problem’s resolution.
        \end{itemize}
        
        The presenter gave the second group some additional disciplines, including:
        \begin{itemize}
            \item Communications – Interpersonal communication is of the utmost importance when something is being explained or presented, mainly if it could negatively impact the organization’s cybersecurity or infrastructure.
            \item Political Science, specifically International Relations – If the individuals who are not adhering to the IT administrators reside in another country or are representatives of another country and are subject to different laws, that could be a complicating factor.
            \item Ethics - To avoid moral and legal pitfalls, it is essential to maintain an ethical approach to conflict resolution—and everything else. One should identify one's desired ethical perspective (utilitarian, deontological, or virtue ethics) and evaluate one's actions and planned actions accordingly.
            \item Law or Legal Studies – If the individuals who are not listening to the IT administrators jeopardize the organization’s cybersecurity (such as a password being stolen because it is written down on a sticky note, which leads to a data breach), there are serious legal consequences for both the organization and the individuals if they do not adhere to the policies and procedures that dictate cybersecurity standards.
        \end{itemize}
        
        For the last group, the presenter added:
        \begin{itemize}
            \item Information Technology (IT) – Though their focus is on cybersecurity, the Incident Response Plan, the Continuity of Operations Plan, and the information systems that allow the organization to operate, defend, and recover from an attack are just as important.
            \item Law or Legal Studies – If an attacker attempts to breach organizational systems to steal information, that is a breach of law. Still, it is also important to understand the potential legal fallout of an attack (such as fines, updating legal documents, or cybersecurity regulation compliance) and whether the attack was successful.
            \item Psychology – In addition to the attackers' reasoning, the defenders' psychology and actions are just as crucial to understanding how one might improve.
            \item Ethics – The defenders must conduct themselves both ethically and within the bounds of the law so as not to suffer for defending their organization. Additionally, suppose the organization or an external security firm performs penetration testing or another simulation for testing digital defenses. In that case, ethics play a crucial part in the rules of engagement for those scenarios.
        \end{itemize}
        
        \subsubsection{THINK: Identify Resources (Including Other Disciplinary Experts)}

        \indent
        
        Next, the presenter asks the participants to identify resources for their problems. Because the workshop participants only consider issues through an exercise, they do not have the specifics for things such as the budget or a timeline for the effort. However, they have experience and knowledge about potential resources that would be useful to their effort. With input and guidance from the presenter, each group develops a list of resources.
    
        The first group (hardware failure) listed the following:
        \begin{itemize}
            \item The IT or infrastructure team monitoring the mission-critical server
            \item The person or team who designed the server’s setup and connections (if they are not the same as the IT or infrastructure team doing the monitoring)
            \item The organization’s Chief Information Officer (CIO)
            \item The server and its components
            \item Backup hardware for the server’s components
            \item The organization’s Backup Plan
            \item The organization’s Disaster Recovery Plan
            \item The organization’s Continuity of Operations Plan
            \item The writer(s) and those in charge of the business plans
            \item The policies and procedures that govern the mission-critical server
            \item Expert contacts outside the organization, such as at a national laboratory
            \item The Internet (Google!)
        \end{itemize}
    
        The second group (people) came up with:
        \begin{itemize}
            \item The IT team
            \item The IT team’s upper management (CIO, etc.)
            \item The troubled individual’s manager
            \item The organization’s Human Resources (HR) department
            \item The organization’s Employee or Workplace Dispute Resolution team
            \item Psychology experts, inside or outside of the organization
            \item Ethics experts, inside or outside of the organization
            \item Change management consultants who can assist with the changes asked of the organization and provide guidance and understanding on how to manage difficult changes and individuals 
            \item The policies and procedures that need to be followed
            \item The writer(s) of the policies and procedures 
            \item The organization’s legal counsel
            \item The organization’s International Relations specialists, if the troubled individual is in a different country
        \end{itemize}

        The third group (defending systems from attackers) identified their resources as:
        \begin{itemize}
            \item The organization’s defenders (IT or security teams)
            \item The organization’s IT infrastructure
            \item The organization’s Chief Information Security Officer (CISO)
            \item The organization’s in-place security controls
            \item The organization’s Incident Response Plan
            \item The organization’s Incident Response Team
            \item External security firms (who can do things like penetration testing)
            \item Internal or external cybersecurity experts
            \item Internal or external criminology experts
            \item Internal or external psychology experts
            \item Internal or external ethics experts
            \item The organization's legal team
            \item The organization’s security policies
            \item Fellow employees who can report suspicious activity 
        \end{itemize}
    
        \subsubsection{PLAN: Consider Alternate Perspectives}

        \indent
        
        The presenter offers an example. Imagine that they (the presenter) are really good at pinball. They think they’re the best pinball player because they’re so good at it. They’ve beaten all their friends at pinball using someone’s home pinball machine, and they always have the highest score. One day, the presenter goes to the local arcade and plays some pinball. They finish their game and expect to be at the top of the scoreboard. It turns out they are tenth on the scoreboard. Beyond being humbled, the presenter realizes they aren’t the best at pinball. They decide they want to improve to beat that top score. What caused that realization? Input (or, in this case, gameplay) from other individuals. People who offered up their game scores for comparison against the presenter’s score.
        
        The presenter then asks the participants: “By a show of hands, who here is biased?”

        Some individuals are reluctant, but the entire room raises their hands.

        The presenter remarks that that’s correct. Every person is biased, regardless of how much they wish to be impartial or unbiased. So then, how does one accurately assess their biases? The first step is acknowledging that one does not know everything. But even with that acknowledgment and some biases being more evident than others, identifying bias can be challenging. That is where the trusted opinion from another person or resource can help. As with the pinball example, realizing that other individuals have something to offer and then hearing their input is essential to improvement. It also allows one to get in the right mindset to consider alternate perspectives. 

        Considering alternate perspectives can be difficult, especially if one needs to gain experience with the process. The presenter asks participants to review their stakeholders and resources with the following questions in mind: What does [the resource] think of this problem? How do they approach it? Is that approach viable in this context? How does it differ from this team’s perspectives? How might that concept be incorporated into a solution?
        
        As participants answer those questions, the presenter asks groups to write down and share the alternate perspectives they come up with.

        The first group (hardware failure) identified the following resources and their perspectives:
        \begin{itemize}
            \item The IT team wants to maintain the critical server's uptime through physical hardware and digital software backups with hot-swappable devices and components.
            \item The CIO’s goals align with the IT team’s goals: to maintain a consistent uptime with easily accessible backups and options to fall back on in the event of a failure.
            \item The organization and business leaders want the critical server to function consistently. When a hardware failure occurs, they want the problem solved quickly so there is minimal impact on business operations.
            \item The organizational Backup Plan, Disaster Recovery Plan, and Continuity of Operations Plan will offer tactics and processes for hardware failure recovery should administrators struggle during a severe outage. If the plans do not provide guidance for hardware failures, those processes should be written into each as soon as possible.
            \item An ethicist would most likely want to notify affected parties as soon as an incident occurs and keep them updated until the issue is resolved, ensuring everyone is treated ethically and respectfully.
        \end{itemize}

        The second group (people) identified the following perspectives:
        \begin{itemize}
            \item The IT team wants to resolve the problem quickly by contacting management to get the troubled person to adhere to the procedure or process they refuse to follow.
            \item The Workplace Dispute Resolution team members want to reach a solution that works for both parties, not just strong-arming the troubled individual into following a policy without understanding it.
            \item The organization wants the issue resolved so it can continue business as usual but does not want to compromise security to achieve this.
            \item An internal or external psychology expert would seek to understand the behaviors of the individuals involved and the motivations behind them to find a solution.
            \item An ethicist would be concerned with the ethics of the situation and, based on their ethical perspective, would want to resolve the dispute ethically.
            \item A change management consultant or consulting team would seek to resolve the dispute by offering strategies for improving business processes and assessing and implementing change management processes to alleviate the strain between employees. Their support would most likely consist of working with middle- and upper-level management to design solutions and processes to benefit the organization for the current and future problems they might encounter regarding people.
            \item The organization’s legal team would seek to ensure all the laws and regulations are being adhered to and that there are no legal repercussions for whatever solution is reached.
        \end{itemize}

        The last group (defending systems from attackers) chose:
        \begin{itemize}
            \item The organization’s defenders seek to minimize the impact of any attack, including putting in prevention, detection, and response measures for security incidents.
            \item The attackers would seek to do the most damage possible, exfiltrate large amounts of data, or lock down data for extortion. Their motivations range from personal vendettas to monetary reasons to nation-state actors seeking to steal information.
            \item An external security firm would seek to assess, identify, and mitigate any vulnerabilities they discover to strengthen the organization's cybersecurity.
            \item A criminology expert would provide insight into how criminals think, behave, are motivated, and use potential attack methods based on their ultimate goals.
            \item An ethicist would strive to ensure that defensive tactics and defenders' responses (hackback, etc.) are ethical and that any penetration testing performed is clearly outlined in writing and within ethical bounds.
            \item A psychology expert would offer knowledge about preventing, finding, understanding, and removing insider threats.
            \item The organization's legal counsel would want all defender operations to be transparently outlined in a legal document to prevent negative repercussions. A professional lawyer could also provide insight, adding to the information from an ethicist and psychologist, to further enhance organizational defense practices and procedures.
            \item Organizational employees may or may not be motivated to report suspicious activity they observe based on their loyalty to the organization, the protections offered to a reporter, and any rewards provided to someone who reports a legitimate threat.
        \end{itemize}

        \subsubsection{PLAN: Plan Goals, Outcomes, and Effects}

        \indent
        
        The presenter asks each group to revisit and describe their efforts' goals, outcomes, and effects. The first group, focusing on hardware failure, states that their goal is to minimize the negative impact on the organization due to the failure. Their intended outcome is to make the organization resilient to hardware failures on that mission-critical asset. The effects include an improved incident response plan, updated procurement procedures, and a resilient workforce capable of responding to severe events. The second group, whose problem was people, aims to have individuals listen to instructions and follow proper procedures to improve the organization’s overall security. Their expected outcomes include improved adherence to policy, organizational security, and a streamlined process for IT team members dealing with additional people problems. Lastly, the third group, whose focus was defending their systems from attackers, aims to successfully defend against an attack and protect their organization’s data. Their intended outcome is to have the organization be unaffected by an attack. The expected effects include lessons learned from the successful defense for improved cybersecurity, increased support for the organization (both externally and internally) because of the successful protection, and a resilient cybersecurity infrastructure that is well-protected from attacks.

        \subsubsection{PLAN: Brainstorm Solutions}

        \indent
        
        Though they have already touched on potential solutions, the presenter asks each group to brainstorm solutions to their problems, including actions or tactics that contribute to their effort’s primary goal. Each group brainstorms potential solutions with the presenter’s feedback and presents their answers to the other participants.

        The first group (hardware failure) devised the following:
        \begin{itemize}
            \item Order and keep parts for the mission-critical server on hand in the event of an emergency.
            \item Develop and maintain a Backup Plan for all critical pieces of the IT infrastructure.
            \item Create a procedure for procurement of mission-critical hardware if one does not already exist.
            \item Consult with legal counsel to keep procedures up to date and help the organization defend from potential litigation.
            \item Identify and know the key individuals who are reliant on the mission-critical server and keep them updated should the hardware fail.
            \item Prepare for and expect an eventual hardware failure on the mission-critical server by planning and implementing redundancy in the IT infrastructure.
            \item Continually seek new methods for improved actions and reactions in the event of a hardware failure.
        \end{itemize}

        The second group (people) identified:
        \begin{itemize}
            \item Seek counsel from the Workplace Dispute Resolution team on what actions should be taken.
            \item Work with the Workplace Dispute Resolution team to resolve the situation.
            \item Talk to the organization’s legal counsel to ascertain any additional legal context of the situation.
            \item Direct the problem person to HR to resolve any grievances.
            \item Develop a curriculum based on the experience, potentially with input from psychology and ethics experts, and then train staff to navigate difficult situations better.
            \item Hire a change management consultant to offer guidance, train staff on change management processes or managing difficult individuals during a change, and design and implement improved organizational processes.
        \end{itemize}

        The last group (defending systems from attackers) came up with:
        \begin{itemize}
            \item Assess current safeguards and controls to identify any gaps in organizational defenses.
            \item Train and maintain a consistent training schedule for defenders in protecting organizational systems.
            \item Perform a penetration test, either internally or with the help of an external security firm, to assess both defender response and organizational defenses.
            \item Review and update the existing Incident Response Plan for tactics and defensive measures during a cyber attack.
            \item Consult a criminology expert for insight on potential attack types, attacker motivations, or attacker behavior to assist defenders in more fully understanding the range of possible cyber attacks.
            \item Work with an ethicist to review and determine the ethical actions and reactions defenders can perform in the event of a cyber attack.
            \item Seek input from a psychology expert to develop training, understanding, and processes to prevent, find, and remove insider threats, particularly those allowing external attackers to break into organizational systems.
            \item Consult an external security firm, a psychological expert, and potentially an ethical expert for guidance on developing or updating a reporting process and developing an organizational culture of camaraderie so that employees will be more likely to report suspicious behavior.
        \end{itemize}
        
        \subsubsection{DO: Execute Action Items}

        \indent
        
        After creating their solutions, the presenter asks participants to choose the ones they believe would benefit the organization the most. This can be all of them or only some of them. Participants should select their solutions and consider what action items need to be performed for their solution to be fully realized. Some solutions already have action items, but some might need more explanation on the actions to be taken.

        The first group (hardware failure) selected their action items as:
        \begin{itemize}
            \item Order and keep parts for the mission-critical server on hand in the event of an emergency.
            \item Develop and maintain a Backup and Restoration Plan for all critical pieces of the IT infrastructure.
            \item Create a procedure for procurement of mission-critical hardware if one does not already exist.
            \item Consult with legal counsel to keep procedures up to date and help the organization defend from potential litigation.
            \item Identify and know the key individuals who are reliant on the mission-critical server and keep them updated should the hardware fail.
            \item Prepare for and expect an eventual hardware failure on the mission-critical server by planning and implementing redundancy in the IT infrastructure.
            \item Continually seek new methods for improved actions and reactions in the event of a hardware failure. They will do this by attending conferences, meetings, and other events that foster the sharing of ideas.
        \end{itemize}

        The second group (people) selected:
        \begin{itemize}
            \item Seek counsel from the Workplace Dispute Resolution team on what actions should be taken.
            \item Work with the Workplace Dispute Resolution team to resolve the situation.
            \item Talk to the organization’s legal counsel to ascertain any additional legal context of the situation. Use the information they provide to assist in making decisions on the next steps.
            \item Develop a curriculum based on the experience, potentially with input from psychology and ethics experts, and then train staff to navigate difficult situations better.
            \item Hire a change management consultant to offer guidance, train staff on change management processes or managing difficult individuals during a change, and design and implement improved organizational processes.
        \end{itemize}

        Finally, the third group (defending systems from attackers) settled on the following action items: 
        \begin{itemize}
            \item Assess current safeguards and controls to identify any gaps in organizational defenses. Use this assessment in conjunction with a risk assessment to determine the most significant threats from outside attackers and insider threats.
            \item Train and maintain a consistent training schedule for defenders on protecting organizational systems and recognizing and addressing insider threats. Seek industry best practices for training curriculum.
            \item Perform a penetration test, either internally or with the help of an external security firm, to assess both defender response and organizational defenses.
            \item Review and update the existing Incident Response Plan for tactics and defensive measures during a cyber attack. Confer with management to ensure the Incident Response Plan is reviewed at least once a year.
            \item Consult a criminology expert for insight on potential attack types, attacker motivations, or attacker behavior to assist defenders in more fully understanding the range of possible cyber attacks. Utilize the criminologist's instruction to train and inform employees about actions they can take to help prevent outside attackers from successfully breaching systems.
            \item Work with an ethicist to review and determine the ethical actions and reactions defenders can perform in the event of a cyber attack. Ensure defenders know the moral and legal obligations for defending and retaliating before, during, and after an attack.
            \item Seek input from a psychology expert to develop training, understanding, and processes to prevent, find, and remove insider threats, particularly those allowing external attackers to break into organizational systems.
            \item Consult an external security firm, a psychological expert, and potentially an ethical expert for guidance on developing or updating a reporting process and developing an organizational culture of camaraderie so that employees will be more likely to report suspicious behavior.
        \end{itemize}

        \subsubsection{DO: Review, Learn, Revise, Repeat}

        \indent
        
        The presenter concludes the workshop by discussing the final step of the framework.
        
        \textbf{Review:} Consistently reviewing policies, procedures, and processes is critical to an organization’s success. Whether new or old processes, review is vital for fostering an environment of continual learning.
        
        \textbf{Learn:} Just as with reviewing, learning is critical for success. With the rapid pace of technology, experts should always try to approach situations with the mindset of finding something new to learn. Learning from a situation allows for improvement and better future actions and reactions.
        
        \textbf{Revise:} Following the learning phase, one should revise whatever one determines needs to be changed. Can one honestly say they learned from something if nothing they do or say changes?
        
        \textbf{Repeat:} This final phrase encourages individuals to continually seek to learn and improve. Cyber threats will continue to evolve, and so should the experts who defend against them. Experts can evolve by embracing new ways of thinking, learning, and defending.

    \subsection{The Need For Multiple Disciplines}

    \indent
    
    The use cases above are just three examples of where the framework might be applied. Nonetheless, it should be repeated that the framework is not the singular focus of this paper; transdisciplinary thinking is. The framework is simply a tool to encourage and assist experts in broadening their thinking and methods. The examples in this paper show that many disciplines outside of the hard sciences can offer additional insight into a cybersecurity scenario, like law, ethics, psychology, or economics. This is especially true when one is confronting an intimidating or formidable issue. However, the need for multiple disciplines is not exclusive to cybersecurity. Transdisciplinary thinking can help experts of all disciplines to rise and meet the complex challenges of the modern world. By working together, learning from one another, and creating innovative solutions, experts can help make the world a better place for everyone.

\section{LIMITATIONS AND FUTURE WORKS}

\indent

This work's most prominent and apparent limitation is that the framework has yet to be applied to a situation or policy by an experienced cybersecurity expert. While theoretical applications help one understand the framework and its uses, the framework’s true efficacy can only be tested by real-world experts who apply it in their unique situations. Input from experts who have tried to apply the framework will be instrumental in refining and moving towards a transdisciplinary way of thinking. Another limitation of this work is its novelty. This paper’s literature review identified a gap in the current literature on transdisciplinary cybersecurity. Because this framework is new, it is subject to the author's bias, limited knowledge, and limited research on which to build. However, because it is novel, there is the potential for future research to test, update, and adapt the framework and its concepts to a variety of different fields. It is anticipated that with practitioner feedback, the framework can be shaped and refined into an even better tool to assist individuals with applying transdisciplinary approaches. 

Future potential research directions for applying transdisciplinary thinking include business, education, and medicine. Businesses are increasingly reliant on and integrated with technology. Not only does business border technology, but disciplines such as law, ethics, or communications are integral parts of a business’ functions. Furthermore, modern education is increasingly moving towards the complete digitization of records. Due to the COVID-19 pandemic, education options moved from programs exclusively delivered in-person to some programs being offered entirely online. Transdisciplinary thinking could be applied to the school systems or planning a transdisciplinary program or degree. In addition to those two potential avenues of research, medicine is an ever-growing field filled with cutting-edge technology and research. Not only does it involve technology, but ethics are of particular importance when preserving human life. A transdisciplinary way of thinking might benefit medical experts, their practice, or the administration of medical services.

Lastly, the U.S. Joint Office of Energy and Transportation, introduced in Section 5.2.2, deserves special mention. This office was created as part of the Bipartisan Infrastructure Law to move the United States towards nationwide electric vehicle adoption \cite{DOE-DOT}. Lawmakers saw the need to merge disciplines and offices for modern Americans' unique challenges. The Joint Office is a prime example of what might be achieved by shifting towards a transdisciplinary perspective for addressing a problem. Like this framework, because it is new, there are bound to be issues, expected and unexpected. However, it is hoped that by leveraging the brilliance of other disciplinary experts, integration and the creation of new methods and solution methods might be achieved.

\section{CONCLUSION}

\indent

Cybersecurity is a challenging but rewarding field. Due to the proliferation of technology, cybersecurity is now more than just computer science or mathematics. The discipline must evolve into one that encourages and incorporates transdisciplinary thinking to equip experts with tools to better defend themselves and those they care for. This paper proposed a framework to enable experts to build upon and enhance their traditional thinking with transdisciplinary thinking. Despite the difficulties facing modern cybersecurity experts, it is hoped that with this framework, experts can grow and improve to tackle increasingly complex problems. The debut of this framework begs for more testing. It may need to be modified, updated, and adapted to suit the needs of disciplinary experts. However, armed with knowledge, tools, and experience, cybersecurity experts can make the world a safer place. Because of technology’s integration with modern life, cybersecurity can no longer be considered an exclusive, homogenous discipline. Cybersecurity, \textit{good} cybersecurity, is transdisciplinary. Further research into the effective integration of transdisciplinarity in cybersecurity will only strengthen the discipline.

\newpage

\printbibliography[title= \hfill REFERENCES \hfill 
\vspace{1pc}]

\newpage

\appendix
\phantomsection
\section*{\centering APPENDIX A}
\addcontentsline{toc}{section}{APPENDIX A.  
     PERMISSIONS}
     \vspace{1pc}
\section*{\centering PERMISSIONS}

\vspace{1pc}

\normalsize

\indent 

To the best of the author's knowledge, no work cited in this document required a request for permission. All figures in this work are original images generated using the free website, \textit{Creately}. Any resemblance to other images or figures is purely coincidental and unintended.

\newpage

\begin{center}
    A TRANSDISCIPLINARY APPROACH TO CYBERSECURITY: \\
    A FRAMEWORK FOR ENCOURAGING \\ TRANSDISCIPLINARY THINKING

    \vspace{0.5cm}
    by
        
    \vspace{0.3cm}
    Emily Kesler

    \vfill

    \end{center}

    \indent Permission to make digital or hard copies of all or part of this work for personal or classroom use is granted without fee provided that copies are not made or distributed for profit or commercial advantage and that copies bear this notice and the full citation on the last page. To copy otherwise, to republish, to post on servers or to redistribute to lists, requires prior specific permission and may require a fee.

    \vfill

    \begin{center}
        \includegraphics[scale=0.6]{figures/NMTLogo.jpg}
    \end{center}

\end{document}